\documentclass[a4paper,11pt]{article}
\usepackage[utf8]{inputenc}
\usepackage[english]{babel}
\usepackage{amsfonts}     
\usepackage[psamsfonts]{amssymb}
\usepackage[bottom]{footmisc}
\usepackage{amsmath}
\usepackage[T1]{fontenc}  
\usepackage{graphicx}
\usepackage[dvipsnames]{xcolor}
\usepackage[final]{pdfpages} 
\numberwithin{equation}{section}

\allowdisplaybreaks

\def\b1{\bar 1}

\def\kbar{{\mathchar'26\mkern-9muk}}
\def\p{{\partial}}
\def\cA{{\cal A}}

\begin{document}

\title{\bf The fuzzy BTZ}

\author{Ilija Buri\'c $^{1}$\thanks{ilija.buric@df.unipi.it}\
                           \ and
        Maja Buri\'c $^{2}$\thanks{majab@ipb.ac.rs}\
        \\[15pt]
        $\strut^{1}${\it Department of Physics, University of Pisa, Largo B. Pontecorvo, 56127 Pisa, Italy}
        \\[5pt]
        $\strut^{2}${\it Faculty of Physics, University of Belgrade, Studentski trg 12, 11001 Belgrade, Serbia}
        }

\date{}

\maketitle

\begin{abstract}
    We introduce a model of a noncommutative BTZ black hole, obtained by quantisation of Poincar\'e coordinates together with a moving frame. The fuzzy BTZ black hole carries a covariant differential calculus, satisfies Einstein's equations and has a constant negative curvature. The construction passes through a larger space, the fuzzy anti-de Sitter, and implements discrete BTZ identifications as conjugations by a unitary operator. We derive the spectrum of the suitably regularised radial coordinate: it consists of a continuum of scattering states outside the horizon $r_+$ and an infinite discrete set of bound states inside.
\end{abstract}

\newpage

\tableofcontents

\section{Introduction}

There are numerous reasons to study geometry and physics of  noncommutative spaces. Besides the purely mathematical interest in geometries that may be associated to general, or to $C^\ast$-algebras,
spaces with noncommuting coordinates appear as effective in string theory and loop quantum gravity. Extending the common quantum-mechanical and general-relativistic intuition, one expects that noncommutative geometry is a general feature of models of spacetime that emerge from quantum  gravity, and that it can give valuable insight into the physics at the Planck scale.

The adjective "fuzzy" (as in fuzzy space)  is often associated with geometries of finite-dimensional matrix algebras. The term was originally introduced by John Madore in relation to his noncommutative frame formalism \cite{Madore:2000aq}, which also describes geometries of algebras with infinite-dimensional irreducible representations: we use it in this, broader sense.
One of the basic ingredients of the  noncommutative frame formalism is a version of the correspondence principle, a rule that to a classical geometry, described by the moving frame $\{\tilde e_\alpha\}$, associates a noncommutative frame $\{e_\alpha\}$,
\begin{equation}\label{correspondence}
    \tilde e^\mu_\alpha =  \tilde e_\alpha \tilde x^\mu \quad \longrightarrow \quad e^\mu_\alpha =[p_\alpha,x^\mu]\ .
\end{equation}
The elements $p_\alpha$ and $x^\mu$ on the right, referred to as momenta and coordinates, belong to a noncommutative algebra. As functions of coordinates, the noncommutative frame coefficients $\, e^\mu_\alpha(x)\,$  in the commutative limit are equal or reduce to $\, \tilde e^\mu_\alpha(\tilde x)\,$.

The best known example of a fuzzy space is the fuzzy sphere \cite{Madore:1991bw}, for which both coordinates and momenta are taken as generators of $SO(3)$ in the $N$-dimensional unitary irreducible representation. The equation of the sphere
\begin{equation}\label{embedding}
    g_{\mu\nu} x^\mu x^\nu = \text{const}
\end{equation}
is then satisfied due to constancy of the quadratic Casimir in the representation. In general, starting from coordinates and momenta, the frame formalism  systematically develops notions of ordinary differential geometry, such as a metric, connection and curvature, in the noncommutative setup. This leads to differential and tensor calculi closely analogous to commutative ones and allows for the construction of field theories over the noncommutative space. However, consistency between the various geometric structures implies that not any set of operators $(p_\alpha,x^\mu)$ leads to a geometry. In fact, consistency conditions are quite constraining and require, among other things, the momenta $p_\alpha$ to satisfy a quadratic algebra.

Most fuzzy geometries that have been constructed are spaces with a high degree of symmetry and include quantum Euclidean spaces \cite{Fiore:1999sd,Cerchiai:2000qu}, the $h$-deformed hyperbolic plane \cite{cho,Madore:1999fi}, complex projective spaces \cite{Balachandran:2001dd,Grosse:2004wm,Dolan:2006tx} etc. Among the examples are also maximally symmetric spaces in $2n$ dimensions, \cite{Jurman:2013ota,Buric:2017yes}. The latter may be classically defined by the embedding \eqref{embedding} inside the flat $(2n+1)$-dimensional space of appropriate signature. The noncommutative analogue of the embedding relation is achieved using the Pauli-Lubanski vector $W^\mu$ of the isometry algebra $\,\mathfrak{so}(p,q) = \text{span}\{M_{\mu\nu}\}$,
\begin{equation}               \label{still}
    W^\mu = \epsilon^{\mu \mu_1\mu_2\dots\mu_{2n}}\,  M_{\mu_1\mu_2}\dots  \  M_{\mu_{2n-1}\mu_{2n}}  \, .
\end{equation}
The element $\,W_\mu W^\mu$ is the highest order Casimir of $\mathfrak{so}(p,q)$ and thus a constant operator in any irreducible representation. This motivates the identification of fuzzy coordinates as operators $\, x^\mu \sim W^\mu\,$ in a UIR of $SO(p,q)$. However, according to \eqref{correspondence}, the noncommutative geometry is defined not only by the choice of coordinates but also momenta; different choices of momenta leading to geometries with interesting physical properties were considered in \cite{Buric:2015wta}. Yet, a pattern similar to \eqref{still} cannot be used  to construct maximally symmetric fuzzy spaces in odd number of dimensions, as the corresponding isometry groups do not admit the Pauli-Lubanski vector.

It is perhaps fair to say that non-local properties of fuzzy spaces have not yet been discussed conclusively. To address this question would, however, be very desirable in the context of gravity where spacetimes of interesting global causal structure, such as black holes, are of particular importance. In the present work, we will make a step in this direction by constructing a model of the fuzzy BTZ black hole equipped with a local differential calculus as well as a suitably defined global structure. A particular special case of the model may be regarded as the fuzzy AdS$_3$ space. While we cannot follow the ideas of  \cite{Madore:1991bw,Buric:2017yes} directly, the fuzzy space will still be defined in terms of the algebra of operators in an irreducible representation of the classical isometry group, $SO(2,2)$. 

Before describing our construction in more details, let us review some of the models of AdS$_3$ and BTZ spaces within other approaches to noncommutative geometry. A widespread idea for construction of noncommutative geometries is to express the gravitational field in terms of a gauge theory. The appeal of this approach is that  gauge symmetries, in the framework of noncommutative field theory, are very well understood. Moreover, there is a mapping (Seiberg-Witten map) between noncommutative and commutative gauge fields given in the form of a series expansion in the noncommutativity parameter (here denoted by $\kbar$). The Seiberg-Witten map can be interpreted as a perturbative expansion in $\kbar$, thus giving the commutative limit of the theory, as well as the leading-order noncommutative corrections. Specifically, in three dimensions the gravitational action can be written as a difference of two Chern-Simons actions, \cite{Achucarro:1986uwr,Witten:1988hc}. A noncommutative generalisation of this description was given in \cite{Banados:2001xw} for the Euclidean signature, and in \cite{Cacciatori:2002gq} for the Lorentzian signature (see also e.g. \cite{Chamseddine:2000si,Chamseddine:2002fd,Dimitrijevic:2014iwa} for related statements in various spacetime dimensions and signatures). Several subsequent papers considered the linear noncommutative corrections to the BTZ geometry \cite{Kim:2007nx,Chang-Young:2008zbi}, and  thermodynamics \cite{Anacleto:2015kca}.

In these works, the noncommutative coordinate algebra carries the Moyal-Weyl product, with one commuting (central) coordinate and two that satisfy the Heisenberg commutation relations (the same approach was followed in \cite{Pinzul:2005ta} except that the harmonic-oscillator basis for the Heisenberg algebra was used, with matrices eventually being truncated to a finite size). A systematic investigation of commutation relations between coordinates was performed in \cite{Bieliavsky:2002ki,Bieliavsky:2004yp,Dolan:2006hv}, where the authors studied families of regular Poisson structures on the classical BTZ black hole background and subsequently quantised them. Our approach differs from the above ones and the algebra underlying the noncommutative space, which contains both coordinates and momenta, is (a quotient of) the operator algebra in an irreducible representation of $SO(2,2)$.

{\bf Summary of results}

The model proposed here may be regarded as a quantisation of the BTZ black hole for several reasons. Much as for the fuzzy sphere, the isometry group of the classical space underlies its quantisation. Our starting point is the operator algebra $\mathcal{A}$ of an irreducible representation of $SO(2,2)$, the isometry group of AdS$_3$. This group is locally isomorphic to $SL(2,\mathbb{R})\times SL(2,\mathbb{R})$ and the representation taken is the tensor product of discrete series representations of the two factor groups. Inside this algebra, coordinates $x^\mu$ and momenta $p_\alpha$ for the fuzzy AdS$_3$ space are  defined, with momenta being three particular elements of $\mathfrak{so}(2,2)$. When $SO(2,2)$ is viewed as the two-dimensional conformal group, these momenta are generators of translations and dilations. The coordinate operators quantise classical Poincar\'e coordinates. The conformal boundary of the quantised AdS$_3$ is defined: it turns out to be flat and commutative.

Differential and tensor calculi over $\mathcal{A}$ are constructed starting from $(p_\alpha,x^\mu)$ in the framework of the frame formalism. They share many properties of their commutative counterparts. For instance, we show that $\mathcal{A}$ satisfies vacuum Einstein's equations with a negative cosmological constant and that it has a constant negative curvature. Further, we construct the Laplace-Beltrami operator and compute its action on arbitrary differential forms.

In addition to the local calculus, that has similarities to other quantum maximally symmetric spaces such as the $h$-deformed hyperbolic plane \cite{Madore:1999fi} or the fuzzy de Sitter space \cite{Buric:2017yes}, we will take into account the global properties of the BTZ black hole. After all, it is only in its global structure that the black hole differs from AdS$_3$. The fuzzy BTZ operator algebra will be obtained from $\cA$ by making discrete identifications of the form $\,x\sim Ux\,U^{-1}$, with $\,x\in\cA$. Here, $U\in\cA$ is a unitary operator chosen in such a way to reproduce the classical action of BTZ identifications on Poincar\'e coordinates, \cite{Banados:1992gq}. 

Having defined the model, we will begin the exploration of its properties by studying the radial coordinate operator. The eigenvalue problem for this operator turns out to be equivalent to the Schr\"odinger equation for a particle in the inverse square potential. Bound and scattering states correspond to the geometric regions inside and outside the outer black hole horizon, respectively. After regularisation, needed to ensure that the radius satisfies $r^2\geq 0$, the spectrum of bound states is shown to be infinite and discrete.

For noncommutative spaces, a possible way to obtain the classical limit is by defining measurements localised at points of the classical spacetime. This is often done using a set of coherent states. Working in the context of fuzzy AdS$_3$, we will follow a method similar to generalised coherent states of \cite{Perelomov:1986tf} and define a set of semi-classical states $|\xi\rangle$ labelled by points of the classical space, that provide one with the notion of local measurements. The construction ensures that expectation values of Poincar\'e coordinates in states $|\xi\rangle$ coincide with the corresponding classical values. Furthermore, computing uncertainties of coordinate operators reveals the classical limit of the fuzzy AdS$_3$: it arises when the spin labels $l=\bar l$ of discrete series representations tend to (minus) infinity.

The paper is organised as follows. In Section 2 we give a brief review of the AdS$_3$ and BTZ geometry, focusing on those points that will play a role in the subsequent quantisation. After recalling basic elements of the frame formalism, Section 3 defines the fuzzy AdS$_3$ operator algebra and develops in detail differential geometry over it. In Section 4, we perform discrete identifications and solve the eigenvalue problem for the radial coordinate. Section 5 contains the construction of semi-classical states for the fuzzy AdS$_3$ and discusses the classical limit. We end in Section 6 with a summary and discussion of the points to be addressed in the future. Three appendices give more details on some of the calculations, and backgrounds on the BTZ geometry and representation theory of $SL(2,\mathbb{R})$.

\section{Anti-de Sitter space and  BTZ black hole}

In this introductory section we review local and global aspects of the three-dimensional anti-de Sitter space and the Ba\~nados-Teitelboim-Zanelli (BTZ) black hole. Our purpose is two-fold: to establish notation and to point to those properties of AdS$_3$ and BTZ spacetimes that will play a prominent role in the quantisation of later sections. The first of these properties is the fact that the BTZ black hole can be covered by an infinite number of Poincar\'e coordinate patches. While the classical geometry does not depend on the the choice of coordinates, the quantisation procedure does and it is the Poincar\'e coordinates that will be used for quantisation. Relations between the Poincar\'e and other types of coordinates will be spelled out in the first subsection. In the second, we will recall the identification of AdS$_3$ with the Lie group $SL(2,\mathbb{R})$. Representation theory of this group will play an important role in subsequent constructions. For the most part, our conventions follow \cite{Banados:1992gq}.

\subsection{Coordinate systems}

The AdS$_3$ space can be defined as the hyperboloid
\begin{equation}
 -v^2-u^2+x^2+y^2= -\ell^2\,,                                             
\end{equation}
$v$, $u$, $x$, $y\in(-\infty,\infty)$, inside the four-dimensional  flat space of signature $(--++)$,
\begin{equation}
 ds^2=-dv^2-du^2+dx^2+dy^2\ .                                          
\end{equation}
It is a solution to three-dimensional vacuum Einstein's equations with a negative cosmological constant, $\Lambda=-1/\ell^2$. The AdS$_3$ space is often represented in the global coordinate system $(\tau,\rho,\theta)$, in which the line element reads
\begin{equation}\label{AdS}
 ds^2 = \ell^2(-\cosh^2\rho\, d\tau^2 + d\rho^2 + \sinh^2\rho\, d\theta^2)\,,
\end{equation}
or in the polar coordinates $(t,r,\theta)$, with the line element
\begin{equation}                                                               
 ds^2 =- \left(\frac{r^2}{\ell^2}+1\right) dt^2 +\frac{1}{\ \dfrac{r^2}{\ell^2}+1\  }\, dr^2 +r^2 d\theta^2\ . 
\end{equation}
Both $\tau$ and $\theta$ are periodic coordinates and thus AdS$_3$ admits closed timelike curves. For this reason, more usually considered is the universal covering space, obtained by "unwrapping" the $\tau$-direction. For this space, denoted $\widetilde{\text{AdS}_3}$, $(\tau,\rho,\theta)$ with $\,\tau\in(-\infty,\infty)$ is a global system of coordinates.

In considerations here, we will mostly use the Poincar\'e coordinates  $\, (\gamma,\beta,z)$,  introduced as 
\begin{equation}                                                               \label{Poincare}
 z=\frac{\ell}{u+x}\, ,\quad \beta = \frac{y}{u+x}\, ,\quad \gamma = -\frac{v}{u+x} \ . 
\end{equation} 
They cover one half of the hyperboloid. The line element in Poincar\'e coordinates is conformally flat,
\begin{equation}\label{Poincare-line-element}
    ds^2 = \frac{\ell^2}{z^2}\, ( -d\gamma^2 + d\beta^2 + dz^2)\ .
\end{equation}
Properties of various coordinate systems and coordinate changes with more details are discussed in Appendix A. 

Like the anti-de~Sitter space, the BTZ black hole is a solution to vacuum Einstein's equations. Its metric has the general form
\begin{equation}\label{BTZ-metric}
 ds^2 = -N^2 dt^2 +\frac{1}{N^2} \, dr^2 + r^2(N^\phi dt + d\phi)^2\ ,\quad \ N^2 = \frac{r^2}{\ell^2} - M +\frac{J^2}{4r^2}\ , \quad N^\phi = -\frac{J}{2r^2}\ ,
\end{equation}
with $\,t\in (-\infty, \infty)$,  $\,r\in (0,\infty)\, $ and  periodic $\phi$,  $\,\phi\sim\phi+2\pi n \,$. The  BTZ space \eqref{BTZ-metric} describes a rotating black hole of mass $M$ and angular momentum $J$: the singularity at $r=0\,$ is one in the causal structure. It is usually assumed that parameters satisfy $\,|J|\leq M \ell\,$; in the remainder we also take $J\geq0$. The BTZ black hole has two horizons, outer at $\,r = r_+$ and inner at $\,r=r_-\,$, with
\begin{equation}
     r_+\pm r_- =\sqrt{M\ell^2\pm J\ell\, }\ .             \label{horizons}
\end{equation}
As a solution to vacuum Einstein's equations in three dimensions, the BTZ black hole is locally isometric to AdS$_3$ and can be obtained as a discrete quotient of a subset $\widetilde{\text{AdS}'_3}$ of its universal cover. The black hole is constructed from the latter by identifying points $X$ under the action of the discrete subgroup of isometries $\,\Gamma\cong\mathbb{Z}$, generated by a certain Killing vector field $\xi'$, \cite{Banados:1992gq}
\begin{equation}
  X\to e^{2\pi n\,\xi'}\,X \,, \quad n\in \mathbb{Z} \ .             \label{discrete}
\end{equation}
The resulting spacetime consists of an infinite number of regions of types I, II and III, separated by inner and outer horizons. The regions admit local Poincar\'e coordinates in which the metric reads \eqref{Poincare-line-element}, and local sets of coordinates $(t,r,\phi)$ with the metric \eqref{BTZ-metric}. Changes between these coordinates in different regions are collected in Appendix A; they provide an infinite set of Poincar\'e patches that cover the BTZ space.
 
An important property for our purposes is that in each of the regions, the radial coordinate $r\,$ is related to $z$, $\gamma$ and $\beta$ in the same way, 
\begin{equation}                                                         \label{radius-BTZ}
     r^2 - r_+^2 = (r_+^2-r_-^2)\, \frac{B(r)}{\ell^2} =  (r_+^2-r_-^2)\,\frac{y^2 - v^2}{\ell^2}  = (r_+^2-r_-^2)\, \frac{\beta^2 - \gamma^2}{z^2}\ .
\end{equation}
In the following, we shall in calculations often  use  $B(r)$  instead of $\,r$,
\begin{equation}
    B(r)=\ell^2\,\frac {r^2-r_+^2}{r_+^2-r_-^2} =  \ell^2\, \frac{\beta^2 -\gamma^2}{z^2}\ .                                                                                   \label{B(r)}
\end{equation}
In any coordinate patch, the discrete transformation \eqref{discrete} acts as $(t,r,\phi)\mapsto(t,r,\phi+2\pi n)$, and on Poincar\'e coordinates by
\begin{equation}\label{identifications-Poincare}
    z \mapsto z \, e^{-\frac{2\pi r_+ n}{\ell}},\ \quad (\beta - \gamma) \mapsto (\beta - \gamma) \,e^{-\frac{2\pi(r_+ + r_-) n}{\ell}},\ \quad (\beta + \gamma) \mapsto (\beta + \gamma)\, e^{-\frac{2\pi(r_+ - r_-)n}{\ell}}\ .
\end{equation}
This concludes our review of coordinate systems used to describe AdS$_3$ and the BTZ black hole.

\subsection{The identification of AdS$_3$ and $SL(2,\mathbb{R})$}

In the remainder of the text, we will often use the fact that AdS$_3$ is isometric to the Lie group $SL(2,\mathbb{R})$; the notation regarding this group and its Lie algebra is given in  Appendix B. Given a point $X$ in $\mathbb{R}^{2,2}$, one can construct the matrix
\begin{equation}
    g(X) = \frac{1}{\ell}\begin{pmatrix} 
    u+x & y+v\\[2pt]
    y-v & u-x\end{pmatrix}\,,
\end{equation}
and the equation of the hyperboloid is equivalent to $\text{det}(g(X))=1$. Moreover, the mapping $X\mapsto g(X)$ carries the AdS$_3$ metric to the bi-invariant metric on $SL(2,\mathbb{R})$. The identification of the two spaces allows to also identify their isometry groups\footnote{For any Lie group $G$, we write $G_e$ to denote its identity connected component. However, to simplify notation, we will usually just write $SO(2,2)$ instead of $SO(2,2)_e$.} 
\begin{equation}                         \label{SO22}
    SO(2,2)_e = \frac{SL(2,\mathbb{R})\times SL(2,\mathbb{R})}{Z(SL(2,\mathbb{R}))} = \frac{SL(2,\mathbb{R})\times SL(2,\mathbb{R})}{\mathbb{Z}_2}\ .
\end{equation}
An element $(g_1,g_2)\in SL(2,\mathbb{R})\times SL(2,\mathbb{R})$ acts on $g(X)$ according to
\begin{equation}\label{action-on-the-group}
    (g_1,g_2)\cdot g(X) = w g_1 w^{-1}\ g(X)\ g_2^{-1}\ .
\end{equation}
The presence of the Weyl inversion $w$, a particular element of $\,SL(2,\mathbb{R})$ defined in Appendix B, in the action is a matter of convention. The last equation leads to the following expressions for the action of elements of the Lie algebra \eqref{sl2R},
\begin{equation}                \label{classical-generators-zbg}
    H + \bar H = - z\partial_z - \beta\partial_\beta - \gamma\partial_\gamma, \quad E_+ + \bar E_+ = \partial_\beta, \quad E_+ - \bar E_+ = \partial_\gamma\ .
\end{equation}
Since spaces AdS$_3$ and $SL(2,\mathbb{R})$ are isometric, so are their universal covers. Identifications that lead to the BTZ black hole take the form $g\sim \rho_L \,g\, \rho_R\,$ from the $\widetilde{SL(2,\mathbb{R})}$ point of view. Here $\rho_{L,R}\,$ are particular elements of the universal covering group whose projections to $SL(2,\mathbb{R})$ are the $2\times2$ matrices, \cite{Carlip:1995qv}
\begin{equation}
    \rho_L = \begin{pmatrix}
    e^{\pi(r_+-r_-)/\ell} & 0 \\  0 & e^{-\pi (r_+ - r_-) /\ell}
    \end{pmatrix},\qquad    \rho_R =\begin{pmatrix}
    e^{\pi(r_++r_-)/\ell} & 0 \\ 0 & e^{-\pi (r_++r_-)/\ell}
    \end{pmatrix}\ .
\end{equation}
As AdS$_3$ is acted on by $SO(2,2)$, the space of functions $L^1(\text{AdS}_3)$ carries the corresponding geometric representation of this group\footnote{We will not be precise about the classes of functions on which the groups act in this work. The interested reader is referred to \cite{Kirillov}.}. The identification $\text{AdS}_3\cong SL(2,\mathbb{R})$ means that $L^1(\text{AdS}_3)$ is the regular representation of $SL(2,\mathbb{R})$ and its decomposition into irreducibles of $SO(2,2)$ (i.e. $SL(2,\mathbb{R})$-bimodules) follows from the Peter-Weyl theorem. Notice that BTZ identifications break the symmetry and the black hole spacetime has only two independent global Killing vectors.

\section{Differential geometry in  Poincar\'e coordinates}

In this section, we will introduce a model of the fuzzy AdS$_3$ space. After reviewing elements of the noncommutative frame formalism in the first subsection, we will define the algebra $\mathcal{A}$ of noncommutative coordinates and momenta in the second. The remainder of the section develops differential geometry on $\mathcal{A}$. We will first construct an algebra of differential forms over $\mathcal{A}$ with a differential that obeys all the usual properties. This underlying structure will be further refined by introducing a metric and a compatible, torsionless connection. The resulting noncommutative geometry satisfies Einstein's equations with a negative cosmological constant and has a constant negative scalar curvature. In the last part of the section, we construct the Laplace-Beltrami operator and its action on arbitrary differential forms. Global properties that distinguish between AdS$_3$ space and the BTZ black hole are considered in the next section.

\subsection{Noncommutative frame formalism}

Noncommutative space is an associative $\ast$-algebra $\cA\,$ generated by hermitian elements,  noncommutative coordinates $x^\mu$, which obey the commutation relations
\begin{equation}\label{J}
    [x^\mu,x^\nu]=i\kbar J^{\mu\nu}(x^\rho)\ .            
\end{equation}
The commutator $J^{\mu\nu}(x^\rho)$ can, in principle, be an arbitrary function of coordinates. The constant of noncommutativity $\kbar$ is assumed to be larger or of the order of magnitude of the Planck length, $\kbar\geq\ell_{Pl}^2$; the very form of \eqref{J} presumes the existence of a commutative or classical limit. We will in fact allow for a slightly more general definition of a noncommutative space, where $\mathcal{A}$ is any $\ast$-algebra and $x^\mu$ some hermitian elements of it.

In classical general relativity, the gravitational field can be described by a moving frame with local components $\,\tilde e^\mu_\alpha$, $\mu$ being a coordinate index and $\alpha$ a frame index. Generalising this formulation, the noncommutative moving frame is defined as a set of inner derivations $e_\alpha$ of $\cA$, specified by momenta $p_\alpha\in\cA$,
\begin{equation}\label{*}
    e_\alpha f = [p_\alpha,f]\, ,     \quad f\in\cA\ .  
\end{equation}
Notice that by contrast, in commutative geometry all inner derivations vanish and elements of a frame are linear combinations of partial derivatives. A priori, no assumption on the number of frame derivations is made and coordinate and frame indices might run over sets of different cardinalities. We do require, however, that commutators of momenta and coordinates can be expressed solely in terms of coordinates,
\begin{equation}\label{frame}
    e^\mu_\alpha \equiv e_\alpha x^\mu =[p_\alpha,x^\mu] \in \langle x^\mu\rangle\ .
\end{equation}
For spaces satisfying \eqref{J}, we obtain a consistency condition
\begin{equation}
    i\kbar \, [p_\alpha,J^{\mu\nu}] = [e^\mu_\alpha,x^\nu] + [x^\mu,e^\nu_\alpha]\,,
\end{equation}
which can be seen as a differential equation for $J^{\mu\nu}$ in terms of $e^\mu_\alpha$. It relates noncommutativity that is the algebraic structure of $\cA$ with its geometry.

Dual to $\{e_\alpha\}$ is the co-frame of 1-forms $\{\theta^\alpha\}$, $\ \theta^\alpha(e_\beta) = \delta^\alpha_\beta$. Co-frame forms are required to commute with functions
\begin{equation}\label{co-frame-forms}
    [\theta^\alpha, f]=0\,, \quad f\in\cA\,,
\end{equation}
and the space of 1-forms $\Omega^1(\cA)$ is freely generated over $\cA\,$ by $\,\{\theta^\alpha\}\,$. Notice that, due to noncommutativity of $\mathcal{A}$, general 1-forms  do not commute with functions. Tensors of arbitrary rank are defined in terms of $\Omega^1(\cA)$ via tensor products and duals. The condition \eqref{co-frame-forms} can be understood as coming from the requirement that local components of the metric tensor are constant,  $\ g(\theta^\alpha\otimes\theta^\beta)=\eta^{\alpha\beta}$.

The differential $d$ on functions is defined by the usual expression
\begin{equation}\label{d}
    df=(e_\alpha f)\, \theta^\alpha\equiv -[\theta,f] \,, \quad f\in\cA\ .
\end{equation}
The 1-form $\theta=-p_\alpha\theta^\alpha$ is sometimes referred to as the Maurer-Cartan form. One requires that $\mathcal{A}$ and $\Omega^1(\mathcal{\mathcal{A}})$ embed into a larger differential graded algebra of all forms, $\Omega^\ast(\mathcal{A})$, to which $d$ can be extended. Linearity and other relations such as $\,d^2=0\,$ turn out to be quite restrictive. One of the most important consequences of the consistency constraints is that momenta must satisfy a quadratic relation,
\begin{equation}\label{quadratic}
    2P^{\alpha\beta}{}_{\gamma\delta}\, p_\alpha p_\beta -F^\alpha{}_{\gamma\delta}\, p_\alpha-K_{\gamma\delta}=0\,,                                   
\end{equation}
where $P^{\alpha\beta}{}_{\gamma\delta}$, $F^\alpha{}_{\gamma\delta}$ and $K_{\gamma\delta}$ are constants, that is, elements of the centre $Z(\cA)$. These structure constants have a natural interpretation in the algebra of forms $\Omega^\ast(\mathcal{A}).$ The $\,P^{\alpha\beta}{}_{\gamma\delta}\,$ are the coefficients used to define the exterior product of 1-forms,
\begin{equation}\label{coefficients-P}
    \theta^\alpha\wedge \theta^\beta \equiv  \theta^\alpha\theta^\beta = P^{\alpha\beta}{}_{\gamma\delta}\, \theta^\gamma\theta^\delta\ .                             
\end{equation}
Coefficients $\,F^\alpha{}_{\beta\gamma}\,$ define the action of the differential on 1-forms,
\begin{equation}
    d\theta^\alpha = -\{\theta,\theta^\alpha\} - \frac12 F^\alpha{}_{\beta\gamma}\theta^\beta \theta^\gamma\ .
\end{equation}
Finally, the $\,K_{\alpha\beta}\,$ measure the failure of the Maurer-Cartan equation, $\, d\theta+\theta^2 = -\frac12 K_{\alpha\beta}\theta^\alpha\theta^\beta\, $.

In cases when momenta form a Lie algebra, relations \eqref{quadratic} simplify. Coefficients $F^\gamma{}_{\alpha\beta}$ coincide with the Lie algebra structure constants,
\begin{equation}
      [p_\alpha,p_\beta] = F^\gamma{}_{\alpha\beta} \,p_\gamma\,,
\end{equation}
the central charges vanish, $\, K_{\alpha\beta} =0\,$, and $\, P^{\alpha\beta}{}_{\gamma\delta}\,$ is the usual antisymmetrisation, $\, 2P^{\alpha\beta}{}_{\gamma\delta}= \delta^\alpha_\gamma \delta^\beta_\delta - \delta^\alpha_\delta \delta^\beta_\gamma\, $. In particular, frame 1-forms $\theta^\alpha$ anticommute. This implies that the structure of the algebra of differential forms $\Omega^*(\cA)$, up to noncommutativity of functions, is the same as in commutative differential geometry.

The differential structure that we discussed may be refined by addition of a metric and a connection\footnote{In the terminology of \cite{Madore:2000aq}, what we refer to as connection is called a {\it linear connection}, and is to be distinguished from the weaker notion of a {\it Yang-Mills connection}.} in analogy with the commutative case, although structural statements such as the existence and uniqueness of the Levi-Civita connection generally do not hold. While the connection and curvature can be defined abstractly, the existence of a frame allows to describe them efficiently using connection 1-forms $\omega^\alpha_{\ \beta}$ and curvature 2-forms $\Omega^\alpha_{\ \beta}$. Starting with the set $\{\omega^\alpha_{\ \beta}\}$, the covariant derivative and the curvature forms are given by
\begin{equation}
    D\theta^\alpha =- \omega^\alpha{}_\beta\otimes
    \theta^\beta = -\omega^\alpha{}_{\gamma\beta}\,\theta^\gamma
    \otimes
    \theta^\beta, \qquad \Omega^\alpha_{\ \beta} = d\omega^\alpha_{\ \beta} + \omega^\alpha_{\ \gamma}\omega^\gamma_{\ \beta} = \frac12 R^\alpha_{\ \beta\gamma\delta}\theta^\gamma\theta^\delta\ .
\end{equation}
These are the same as classical expressions, except that obviously $\omega^\alpha_{\ \gamma\beta}$ and $R^\alpha_{\ \beta\gamma\delta}$ are elements of the algebra $\mathcal{A}$. For many more details, we refer the reader to \cite{Madore:2000aq}.

In the commutative limit, the formalism is invariant under coordinate changes. Indeed, consider a set of coordinate operators given by $\, x^{\prime \mu} =x^{\prime \mu}(x^\nu)$, with the momenta $p_\alpha$ being fixed. For the frame components one finds by the Leibniz rule
\begin{equation}\label{noncommutative-tensors}
     e^{\prime\mu}_\alpha =[p_\alpha, x^{\prime\mu}] = \frac{\partial x^{\prime\mu}}{\partial x^\nu}\, [p_\alpha, x^\nu]+ O(\kbar)\ .
\end{equation}
We have used that the commutator of coordinates is of order $\kbar$, \eqref{J}. Therefore, in the commutative limit $\kbar\to0$, we have that $e^{\prime\mu}_\alpha$ and $g^{\prime\mu\nu}$ transform as tensors
\begin{equation}
e^{\prime\mu}_\alpha = \frac{\partial x^{\prime\mu}}{\partial x^\nu}\, e^\nu_\alpha\, , \qquad g^{\prime\mu\nu} = \frac {\partial x^{\prime\mu}}{\partial x^\rho}\, \frac{\partial x^{\prime \nu}}{\partial g^\sigma}\, g^{\rho\sigma}\ .
\end{equation}
Outside the limit, \eqref{noncommutative-tensors} shows that the transformation behaviour receives corrections due to noncommutativity. Consequently, it is reasonable to identify two fuzzy spaces with coordinates $x^\mu$ and $x^{\prime\mu}$ and the same momenta $p_\alpha$.

\subsection{Coordinates and frame relations}

To quantise a classical spacetime, one replaces its algebra of functions $\tilde\cA\,$ by a noncommutative algebra $\cA\,$ in such a way that the latter resembles the former in some suitable sense. In the noncommutative frame formalism, the relations \eqref{frame} assume the role of the correspondence principle. A set of coordinate functions $\tilde x^\mu$ and vector fields $\tilde e_\alpha$ that obey $\tilde e_\alpha\tilde x^\mu = \tilde e^\mu_\alpha(\tilde x)\,$ is to be replaced by operators $x^\mu$ and $p_\alpha$ with $\, e_\alpha x^\mu=[p_\alpha,x^\mu] = e^\mu_\alpha(x)$, at least in the leading order. With such a starting point, the differential geometry described in the previous subsection typically shares many properties of the commutative one. When the classical metric depends on just one of the coordinates, it is often possible to obtain a noncommutative frame identical to the commutative one: then $\cA\,$ acquires "the same metric" as $ \tilde \cA\,$.\footnote{By a slight abuse of notation, all quantities from ordinary geometry will carry a tilde from now on. For example, a local Poincare coordinate that was denoted by $z$ in Section 2 will now be written as $\tilde z$.}

Another aspect of the relation between $\mathcal{\tilde A}\,$ and $\mathcal{A}\,$ is the notion of a commutative limit. For example, in the strict deformation quantisation, $\mathcal{\tilde A}\,$ and $\mathcal{A}\,$ are isomorphic as vector spaces and only differ in their algebra structures. Similarly, in the case of the fuzzy sphere, the algebra of functions $\cA_N$ defined in the $N$-dimensional UIR of $\mathfrak{so}(3)$ "tends to" the algebra of functions on the commutative two-sphere as $\,N\to \infty\,$. For us, the commutative limit will be realised as the set of relations \eqref{J} between appropriately chosen coordinate functions.

Following the idea that spacetime symmetries provide a natural framework for quantisation, in the case of  AdS$_3$ and BTZ spaces we will construct $\mathcal{A}$ using the Lie algebra $\,\mathfrak{so}(2,2)$. As the group $SO(2,2)$ is locally isomorphic to $SL(2,\mathbb{R})\times SL(2,\mathbb{R})$, $\,\mathfrak{so}(2,2)$ is a direct sum of two $\,\mathfrak{sl}(2,\mathbb{R})$ subalgebras. Their "antihermitian" generators are denoted by $H$, $E_+$, $E_-$ and $\bar H$, $\bar E_+$, $\bar E_-$, and have the non-zero brackets 
\begin{eqnarray}
 &   [H, E_+] = E_+, \qquad [H, E_-] = - E_-, \qquad [E_+,E_-] = 2H\,,            \label{sl2R}
 \\[4pt]
 &   [\bar H, \bar E_+] = \bar E_+, \qquad [\bar H, \bar E_-] = - \bar E_-, \qquad [\bar E_+,\bar E_-] = 2\bar H\ .
\end{eqnarray}
 Since the frame components $e^\mu_\alpha$ are given by commutators \eqref{*} in the  formalism, we need to identify the spacetime coordinates and momenta simultaneously. It is beneficial to use coordinates that give the simplest form of the metric: we therefore  quantise the AdS$_3$ and BTZ spaces in Poincar\'e coordinates. The corresponding classical orthonormal frame  is 
\begin{equation}
    \tilde e^\mu_\alpha = \frac{\tilde z}{\ell}\, \delta^\mu_\alpha\ .
\end{equation}
This frame is quantised by a set of operators $\,\{z,\beta,\gamma,p_z,p_\beta,p_\gamma\}\,$ that obey relations
\begin{equation}\label{momenta-coords}
    [p_\gamma, \gamma] = \frac z \ell \, ,\qquad [p_\beta, \beta] = \frac z \ell\, , \qquad [p_z,z] =\frac z \ell \,,
\end{equation}
while all other momentum-coordinate commutators vanish. It can be readily verified that these frame relations are satisfied by operators\footnote{Coordinates and momenta may be multiplied by arbitrary powers of the dimensionless quantity $\kbar/\ell^2$. For simplicity we put $\kbar/\ell^2=1$. In the following, in most formulas we will put $\ell=1$ as well.}
\begin{align}
    & p_z = \ell^{-1}\, \,(H + \bar H) , && z = 2 i \, E_+^a \, \bar E_+^{1-a}\,,          \label{Z1} \\[6pt]
    & p_\beta =  \ell^{-1}\, \,(E_+ + \bar E_+), &&\beta = -i E_+^{a-1}\bar E_+^{1-a}\left(H+\frac{a-1}{2}\right) - i E_+^a \, \bar E_+^{-a}\left(\bar H - \frac{a}{2}\right)\,,    \label{B1}    \\[4pt]
    & p_\gamma =  \ell^{-1}\, \,(E_+ - \bar E_+), &&\gamma = -i E_+^{a-1}\bar E_+^{1-a}\left(H+\frac{a-1}{2}\right) + i E_+^a \, \bar E_+^{-a}\left(\bar H - \frac{a}{2}\right)\ .  \label{G1}
\end{align}
For the moment $\, a$ is any positive real number, and elements of $\mathfrak{so}(2,2)$ are regarded as operators acting in the tensor product of two discrete series representations $\,\mathcal{H}=T_l^-$ and $\mathcal{\bar H} = T_{\bar l}^-$ of $\mathfrak{sl}(2,\mathbb{R})$. Therefore, the momentum-coordinate operator algebra is $\,\mathcal{A} = \text{End}(\mathcal{H}\otimes\mathcal{\bar H})$.  Operators $\,iE_+$ and $\,i\bar E_+$ in discrete series representations are positive, so their powers appearing above are well-defined. As will be seen in the next section, unless $a=0,1$, the full algebra $\mathcal{A}$ is generated solely by coordinates (and their inverses), so in particular, frame derivations are inner. Equations (\ref{Z1}-\ref{G1}) define our noncommutative AdS$_3$ space. 

The remainder of this section and the next one are devoted to the study of the associated geometry. To begin, observe that momentum operators form a Lie algebra
\begin{equation}\label{momenta-momenta}
    [p_z,p_\gamma] = \frac 1 \ell \,p_\gamma, \qquad [p_z,p_\beta]=\frac 1 \ell \, p_\beta, \qquad [p_\beta,p_\gamma]=0\ .
\end{equation}
Comparing with \eqref{classical-generators-zbg}, we see that $p_\beta$ and $p_\gamma$ canonically quantise momenta associated with coordinates $\beta$ and $\gamma$. Upon identifying $\mathfrak{so}(2,2)$ with the $(1+1)$-dimensional conformal algebra, $p_\beta$ and $p_\gamma$ are seen to become translation operators, while $p_z$ becomes the generator of dilations. Therefore, our momenta are analogous to ones introduced in \cite{Buric:2015wta} for the four-dimensional de Sitter space.

The classical solution to frame relations \eqref{momenta-coords} consists of commutative coordinates $\,\tilde z,\ \tilde\beta,\ \tilde\gamma$, regarded as multiplication operators on $C^\infty(\text{AdS}_3)$ and frame vector fields $\,\tilde p_z = \tilde e_z,\ \tilde p_\beta = \tilde e_\beta,\ \tilde p_\gamma = \tilde e_\gamma$. Since the momenta defined above are elements of $\mathfrak{so}(2,2)$, they can also be naturally represented as vector fields on AdS$_3$. One thus may wonder whether these vector fields coincide with the classical frame, i.e. whether $\,p_\alpha = \tilde p_\alpha$. One readily sees that they do not: indeed $\tilde p_\alpha$ are not Killing vectors, while $p_\alpha$ by definition are. The relation between $\tilde p_\alpha$ and $p_\alpha$ becomes clearer by considering the (local) foliation of the space into surfaces of constant $\tilde z$. From the form of the metric in Poincar\'e coordinates, it is clear that these surfaces are flat. The foliation is preserved by the map
\begin{equation}\label{beta'}
    \Phi : \text{AdS}_3 \to \text{AdS}_3, \qquad (\tilde z,\tilde\beta,\tilde\gamma) \mapsto (\tilde z',\tilde\beta',\tilde\gamma') = \left(\frac{1}{\tilde z},\frac{\tilde\beta}{\tilde z},\frac{\tilde\gamma}{\tilde z}\right)\,,
\end{equation}
which reduces to the identity on the plane $\tilde z=1$. It is the map $\Phi\,$ that relates $p_\alpha$ and $\tilde p_\alpha$ through the push-forward\footnote{The proof is exhibited by the relation
\begin{equation*}
    - \tilde z\partial_{\tilde z} - \tilde\beta\partial_{\tilde\beta} - \tilde\gamma\partial_{\tilde\gamma} = \tilde z'\partial_{\tilde z'},\quad \partial_{\tilde\beta} = \tilde z' \partial_{\tilde\beta'}, \quad \partial_{\tilde\gamma} = \tilde z' \partial_{\tilde\gamma'}\ . 
\end{equation*}
Vector fields on the left are $\,p_z$, $p_\beta$, $p_\gamma$ and on the right are $\,\tilde p_{\tilde z'}$, $\tilde p_{\tilde\beta'}$, $\tilde p_{\tilde\gamma'}$.}
\begin{equation}\label{palpha-ealpha}
    \Phi^\ast(p_\alpha) = \tilde p_\alpha\ .
\end{equation}
It may be observed that the coordinates $\beta'$, $\gamma'$ and momenta $p_\beta$, $p_\gamma$ form a pair of mutually commuting Heisenberg algebras,
 \begin{equation}\label{boundary}
     \left[\frac \ell 2\, (p_\beta+p_\gamma), \,\frac{\beta+\gamma}{z}\right]=1\, , \qquad    \left[\frac \ell 2\, (p_\beta-p_\gamma),\, \frac{\beta-\gamma}{z}\right]=1\,,
\end{equation}
where $\, \beta'\pm \gamma'=(\beta\pm\gamma)/z\,$ are defined as symmetrised products, \eqref{symmetrised}. These relations can be given an interesting interpretation in terms of the conformal boundary, $\{\tilde z = 0\}$. We have already mentioned that $p_\beta$ and $p_\gamma$ are translation generators of the conformal group of the boundary. As for the coordinates, the metric on any surface $\tilde z =$const is given by
\begin{equation}
    ds^2 = \ell^2\left(d\tilde\beta'^2 - d\tilde\gamma'^2\right)\ .
\end{equation}
Therefore, we may view $\, (\beta\pm\gamma)/z\,$ as lightcone coordinates on the quantum boundary, $\{ z = 0\}$: relations \eqref{boundary} state that the quantum boundary is a commutative flat plane.

\subsection{Differential forms, metric, connection and curvature}

The algebra of differential forms $\Omega^\ast(\mathcal{A})$ that we will use is based on a frame with three elements $\{\theta^z,\theta^\beta,\theta^\gamma\}$. Elements of the frame commute with $\mathcal{A}$ and anticommute with one another. Therefore, we have
\begin{equation}
    \Omega^\ast(\mathcal{A}) = \mathcal{A} \otimes \Lambda_3\,,
\end{equation}
where $\Lambda_3$ denotes the Grassmann algebra on three generators. In accord with the general theory, the differential is defined on functions by
\begin{equation}
    d = \theta^z \text{ad}_{p_z} + \theta^\beta \text{ad}_{p_\beta} + \theta^\gamma \text{ad}_{p_\gamma}\,,
\end{equation}
and on elements of the frame by
\begin{equation}
    d\theta^z=0, \quad d\theta^\beta=-\theta^z\theta^\beta, \quad d\theta^\gamma=-\theta^z\theta^\gamma\ .
\end{equation}
These requirements together with the Leibniz rule define $d$ uniquely and one can explicitly check that $d^2=0$. Thus, we get a consistent differential graded algebra $(\Omega^\ast(\mathcal{A}),d)$. For future convenience, let us spell out the action of $d$ on coordinates, momenta and 2-forms:
\begin{align}
    & dz = z\theta^z, \quad d\beta = z\theta^\beta, \quad d\gamma = z\theta^\gamma\,,\\[4pt]
    & dp_z = -p_\beta \theta^\beta - p_\gamma \theta^\gamma, \quad dp_\beta = p_\beta \theta^z, \quad dp_\gamma = p_\gamma \theta^z\,,\\[4pt]
    & d(\theta^z\theta^\beta)=0, \quad d(\theta^z\theta^\gamma)=0, \quad d(\theta^\beta\theta^\gamma) = -2\theta^z\theta^\beta\theta^\gamma\ .
\end{align}
The metric is given by the same expression as in the commutative case
\begin{equation}\label{metr}
    g^{zz}=1, \quad g^{\beta\beta}=1, \quad g^{\gamma\gamma} = -1\,,          
\end{equation}
so the commutative limit of the metric is correct, in fact exact. Notice that the indices in \eqref{metr} refer to elements of the frame, e.g. $g^{zz}$ stands for $g(\theta^z \otimes \theta^z)$ rather than $g(dz,dz)$. While in general the construction of a Levi-Civita connection $D:\Omega^\ast(\cA)\to\Omega^\ast(\cA)\otimes_{\cA}\Omega^1(\cA)$ requires one to introduce a generalised flip which controls the right Leibniz rule, in the case at hand no such flip is needed. We will set
\begin{equation}\label{connection}
    D\theta^z = -\theta^\beta\otimes\theta^\beta + \theta^\gamma\otimes\theta^\gamma, \quad D\theta^\beta = \theta^\beta\otimes\theta^z, \quad D\theta^\gamma = \theta^\gamma\otimes\theta^z\,,
\end{equation}
and require $D$ to satisfy the ordinary Leibniz rule both from the left and the right. From the action of $D$ one reads off the non-vanishing connection 1-forms
\begin{equation}
    \omega^z_{\ \beta}=\theta^\beta, \quad \omega^z_{\ \gamma} = -\theta^\gamma, \quad \omega^\beta_{\ z} = -\theta^\beta, \quad \omega^\gamma_{\ z} = -\theta^\gamma\ .
\end{equation}
To show that $D$ is torsionless, it is enough to verify that the torsion $\Theta$ vanishes on elements of the frame. Recall that $\Theta$ is the map from 1-forms to 2-forms given by $\,\Theta = d - \pi\circ D\,$, where $\,\pi(\theta^\alpha\otimes\theta^\beta)=\theta^\alpha\theta^\beta$. We have
\begin{equation}
    \Theta(\theta^z) = \theta^\beta \theta^\beta - \theta^\gamma\theta^\gamma =0, \quad \Theta(\theta^\beta) = -\theta^z\theta^\beta-\theta^\beta\theta^z=0, \quad \Theta(\theta^\gamma) = -\theta^z\theta^\gamma-\theta^\gamma\theta^z=0\ .
\end{equation}
The compatibility of $D$ with the metric is expressed by the equation
\begin{equation}
  \omega^\alpha{}_{\eta\zeta}\, g^{\delta\zeta} + \omega^\delta{}_{\eta\zeta}\, g^{\alpha\zeta} = 0\ .
\end{equation}
One can verify the last equation by substituting the connection 1-forms. From the connection one also constructs the curvature tensor. Curvature 2-forms are found to be
\begin{equation}
    \Omega^z_{\ \gamma} = \theta^z\theta^\gamma,\quad \Omega^z_{\ \beta} = -\theta^z\theta^\beta, \quad \Omega^\beta_{\ z} = \theta^z\theta^\beta, \quad \Omega^\beta_{\ \gamma} = \theta^\beta\theta^\gamma, \quad \Omega^\gamma_{\ z} = \theta^z\theta^\gamma, \quad \Omega^\gamma_{\ \beta} = \theta^\beta\theta^\gamma\,,
\end{equation}
and they lead to the following set of non-zero components of the Riemann tensor
\begin{align}
   & R^z_{\ \beta z\beta} = - R^z_{\ \beta\beta z} = -1, \quad R^z_{\ \gamma z \gamma} = -R^z_{\ \gamma\gamma z} = 1, \quad R^\beta_{\ z\beta z} = -R^\beta_{\ zz\beta} = -1\,,\\
   & R^\gamma_{\ z\gamma z} = -R^\gamma_{\ zz\gamma} = -1, \quad R^\beta_{\ \gamma\beta\gamma}=-R^\beta_{\ \gamma\gamma\beta}= 1, \quad R^\gamma_{\ \beta\gamma\beta} = -R^\gamma_{\ \beta\beta\gamma} = -1\ .
\end{align}
These values coincide with classical expressions. Therefore, the components of the Ricci tensor do so as well,
\begin{equation}
    R_{zz} = -2, \quad R_{\beta\beta} = -2, \quad R_{\gamma\gamma} = 2\ .
\end{equation}
In particular, Einstein's equations $\, R_{ab}=-2g_{ab}\, $ are satisfied.

\subsection{Laplace-Beltrami operator}

As in commutative geometry, the Riemannian Laplace-Beltrami operator may be constructed from the differential and the Hodge star operation. Since the metric components of the geometry discussed above are the same as those of the commutative AdS$_3$, so is the Hodge operator
\begin{align}
    & \ast1 = \theta^z \theta^\beta \theta^\gamma, \quad \ast \theta^z = \theta^\beta \theta^\gamma, \quad \ast \theta^\beta = \theta^\gamma \theta^z, \quad  \ast\theta^\gamma = \theta^\beta \theta^z\,,\\[4pt]
    & \ast(\theta^z\theta^\beta) = \theta^\gamma, \quad \ast(\theta^z \theta^\gamma) = \theta^\beta, \quad \ast(\theta^\beta\theta^\gamma) = -\theta^z, \quad \ast(\theta^z\theta^\beta\theta^\gamma)=-1\ .
\end{align}
The $\ast$ is an $\mathcal{A}$-left-right linear map $\Omega^\ast(\mathcal{A})\to\Omega^\ast(\mathcal{A})$ and satisfies $\ast^2=-1$. The simplest way to define the Laplacian from $d$ and $\ast$ passes through the co-differential $\delta$. On $p$-forms the co-differential is defined by $\delta = (-1)^{p-1}\ast d \ast$. The Laplacian then reads
\begin{equation}\label{Laplacian}
    \Delta = d\delta + \delta d\ .
\end{equation}
Results of the previous subsection allow us to find the action of $\Delta$ on arbitrary forms. For functions, a computation gives
\begin{equation}\label{laplacian-functions}
    \Delta f = -[p_z,[p_z,f]] - [p_\beta,[p_\beta,f]] + [p_\gamma,[p_\gamma,f]] + 2 [p_z,f]\ .
\end{equation}
Similarly for 1-forms, we get
\begin{align}\label{lap-1forms}
    &\Delta(f_z\theta^z + f_\beta\theta^\beta + f_\gamma\theta^\gamma) = (\Delta f_z - [p_\beta,f_\beta] + [p_\gamma,f_\gamma])\,\theta^z\\[2pt]
    & + (\Delta f_\beta - 2[p_z,f_\beta] + 3[p_\beta,f_z])\,\theta^\beta + (\Delta f_\gamma - 2[p_z,f_\gamma] + 3[p_\gamma,f_z])\,\theta^\gamma\ .\nonumber
\end{align}
For 2- and 3-forms, the simplest way to find $\Delta$ is by using the property $\ast\Delta = \Delta\ast$. Thus for 2-forms
\begin{align}\label{lap-2forms}
    &\Delta(f_{z\beta}\theta^z \theta^\beta + f_{z\gamma}\theta^z \theta^\gamma + f_{\beta\gamma}\theta^\beta\theta^\gamma) = (\Delta f_{\beta\gamma}+[p_\beta,f_{z\gamma}]-[p_\gamma,f_{z\beta}])\,\theta^\beta\theta^\gamma\\[2pt]
    &+ (\Delta f_{z\beta}-2[p_z,f_{z\beta}]-3[p_\gamma,f_{\beta\gamma}])\,\theta^z \theta^\beta + (\Delta f_{z\gamma}-2[p_z,f_{z\gamma}]-3[p_\beta,f_{\beta\gamma}])\,\theta^z\theta^\gamma\ .\nonumber
\end{align}
The action on 3-forms is written directly from \eqref{laplacian-functions}
\begin{equation}\label{lap-3forms}
    \Delta(f_{z\beta\gamma}\theta^z\theta^\beta\theta^\gamma) = (\Delta f_{z\beta\gamma})\theta^z\theta^\beta\theta^\gamma\ .
\end{equation}
With this we conclude the discussion of the local differential geometry over the noncommutative AdS$_3$ space $\mathcal{A}$.

\section{Discrete quotient and spectra}

Differential geometry that was developed in the previous section is very general in the sense that it only depends on commutation relations between momenta, and those between momenta and coordinates. Therefore, the geometry can be constructed over any algebra $\mathcal{A}$ which contains elements obeying these relations. Moreover, since the relations in a frame are local, one cannot use them to distinguish between locally isometric spaces such as AdS$_3$ and the BTZ black hole.

In this section, we turn to global properties and find the following. It is possible, in the proposed noncommutative model, to implement the action of discrete identifications on Poincar\'e coordinates via conjugation by a unitary operator $U$ and thereby obtain the fuzzy BTZ black hole. In order to do this, the parameter $a$ in (\ref{Z1}-\ref{G1}) is a fixed function of BTZ horizon radii $r_\pm\,$.

We will find the operator $U$ in the first subsection. In the second, the algebra before and after identifications is realised in terms of differential operators acting on a particular function space. The last subsection discusses the operator of radius $r$, or more precisely $B(r)$, \eqref{B(r)}. This operator turns out to be equivalent to a one-particle Schr\"odinger operator with the inverse square potential. Scattering states correspond to the black hole exterior $r>r_+$, and bound states to the interior. The quantum-mechanical Hamiltonian requires regularisation, as it is self-adjoint only formally; the need of regularisation also arises naturally from the physical requirement that $r^2\geq 0$. The regularised Hamiltonian has a continuum of scattering states together with an infinite discrete set of bound states.

\subsection{Discrete identifications}

In \eqref{identifications-Poincare} we have written the effect of discrete identifications that characterise a BTZ black hole on the Poincar\'e coordinates. In the quantum theory, we have two distinct natural possibilities for implementing this action: either to consider the action of the identification group element $(\rho_L,\rho_R)$ in the algebra $\text{End}(\mathcal{H}\otimes\mathcal{\bar H})$ and impose appropriate invariance under it, or to impose invariance under some transformation that reproduces \eqref{identifications-Poincare} on the quantum level. We will follow the second strategy.

Let $z$, $\beta$ and $\gamma$ be the operators (\ref{Z1}-\ref{G1}). One readily verifies the commutation relations
\begin{equation}
   [H - \bar H,z] = (2a-1)z,\quad [H -\bar H,\beta + \gamma] = 2(a-1)(\beta+\gamma), \quad [H-\bar H,\beta-\gamma] = 2a(\beta-\gamma)\ .
\end{equation}
Therefore, these combinations of coordinates are rescaled by finite transformations as
\begin{equation}                   \label{finite-transformations}
   U z U^{-1} = e^{\alpha(2a-1)}z, \quad U (\beta+\gamma) U^{-1} = e^{2\alpha(a-1)}(\beta+\gamma), \quad U (\beta-\gamma) U^{-1} = e^{2\alpha a}(\beta-\gamma)\ ,
\end{equation}
where we have introduced the unitary operator
\begin{equation}\label{U}   
 U = e^{\alpha(H - \bar H)}\ .
\end{equation}
Transformations \eqref{finite-transformations} assume the same form as classical BTZ identifications \eqref{identifications-Poincare}. By demanding that the two coincide, we get an overdetermined set of equations for $\alpha$ and $a$, which however has the unique solution
\begin{equation}\label{aalpha}
    \alpha =- \frac{2\pi r_-}{\ell}, \quad a = \frac{r_+ + r_-}{2r_-}\ .
\end{equation}
In particular, the identification condition fixes the choice of $\,a$ in definition of coordinates (\ref{Z1}-\ref{G1}), which was up to this point arbitrary. In the remainder of the text, $a$ and $\alpha$ will always be assumed to take the values \eqref{aalpha}.

\subsection{Realisation on a function space}

We turn to a concrete realisation of operators (\ref{Z1}-\ref{G1}) on a function space. This will allow us to determine various properties of physically relevant coordinates, such as their spectra or eigenfunctions.

As mentioned above, unitary irreducible representations of the AdS$_3$ isometry group are of the form $\mathcal{H} \otimes \mathcal{\bar H}$, where $\mathcal{H}$ and $\mathcal{\bar H}$ are unitary irreducibles of $SL(2,\mathbb{R})$. We will take these two representations in the "negative" discrete series
\begin{equation}
    \mathcal{H}\cong T^-_l\, , \quad \mathcal{\bar H}\cong T^-_{\bar l}\ ,
\end{equation}
with $\,2l$, $\,2\bar l$ negative integers. Our conventions about these representations are collected in Appendix B. The coordinate-momentum algebra is that of operators on $\,\mathcal{H}\otimes\mathcal{\bar H}$,
\begin{equation}       \label{coord-momentum-algebra}
    \mathcal{A} = \text{End}(\mathcal{H}\otimes\mathcal{\bar H})\ .
\end{equation}
Discrete series representations of $SL(2,\mathbb{R})$ are most commonly realised on the space of holomorphic functions on the upper half-plane (or the Poincar\'e disc). For our purposes, it turns out to be more convenient to work with the Fourier-space realisation in which the generators take the form \eqref{generators-Fourier}, \cite{Vilenkin}. This realisation is based on the fact that holomorphic functions in the upper half-plane are Fourier transforms of functions defined on the positive real line, and makes it easy to take non-integer powers that appear in equations (\ref{Z1}-\ref{G1}). We find
\begin{align}                                                       \label{Fourier}
    & p_z  = x\partial_x + \bar x  \partial_{\bar x} + l + \bar l + 2, && z =2 x^a \, \bar x^{1-a},\\[2pt]
    & p_\beta  =- i(x+\bar x), && \beta+\gamma = -2i\left(\frac{x}{\bar x}\right)^{a-1}\left(x\partial_x + l + \frac{a+1}{2}\right),\label{Fourier2}\\
    & p_\gamma = -i(x-\bar x), && \beta-\gamma = -2i\left(\frac{x}{\bar x}\right)^a\left(\bar x \partial_{\bar x}  + \bar l + 1 - \frac{a}{2}\right)\ . \label{Fourier3}
\end{align}
These operators act on functions of two real variables $\, x,\bar x>0$. The inner product for discrete series representations is written in \eqref{inner-produt-Fourier}: coordinates are hermitian with respect to the inner product on $\mathcal{H}\otimes\mathcal{\bar H}$, while momenta are anti-hermitian. As mentioned above, for $a\neq0,1$, $\mathcal{A}$ is generated by coordinates. Indeed $x$ and $\bar x$ can be written as 
\begin{equation}
    [z^{-1},\beta+\gamma]^{-1} = -\frac{p_\beta + p_\gamma}{2a} = \frac{ix}{a}, \qquad [z^{-1},\beta-\gamma]^{-1} = \frac{p_\beta-p_\gamma}{2(a-1)}= \frac{i\bar x}{1-a}\,,
\end{equation}
while $\partial_x$ and $\partial_{\bar x}$ are obtained from expressions \eqref{generators-of-the-boundary} below.

In the given representation, the discrete element $U$ acts on functions of coordinates $f(x,\bar x)$ as
\begin{equation}
    (Uf)(x,\bar x) = e^{\alpha(l - \bar l)} f\left(e^\alpha x,e^{-\alpha}\bar x\right)\ .
\end{equation}
The form of this action suggests to change variables. We introduce  $\,\chi\in(0,\infty)$ and $\,\eta\in(-\infty,\infty)$ by
\begin{equation}\label{change}
    x = \chi e^\eta, \qquad \bar x = \chi e^{-\eta}\ .                    
\end{equation}
For $\, l=\bar l$, we find that $U$ acts only on $\eta\,$, as a finite translation
\begin{equation}
   H - \bar H = x\partial_x - \bar x \partial_{\bar x} = \partial_\eta\, ,\qquad (Uf)(\chi,\eta) = f(\chi,\eta+\alpha)\ .
\end{equation}
We will require invariance of the BTZ wave functions under the discrete subgroup generated by $U$ by assuming that $\eta$ is a periodic coordinate, 
\begin{equation}\label{eta}
    \eta\sim\eta+ \alpha  n, \quad n\in\mathbb{Z}\ .           
\end{equation}
This restriction is quite similar to the restriction \eqref{phi} imposed on the classical BTZ space, that coordinate $\phi$ be periodic. After identifications, one is supposed to quantise only those functions which are invariant under \eqref{identifications-Poincare} on the classical level, that is,  functions of $\,(t,r,e^{i\phi})$.

To be more explicit, let us look at a concrete case. For definiteness, consider the black hole with $r_+ = 2 r_-$. Then a simple example of a classically invariant coordinate is $(\tilde\beta+\tilde\gamma)^2/\tilde z$. To quantise this coordinate, one has to choose operator ordering. We do not commit to a general ordering prescription in this work, but for the sake of the argument, let us take the following symmetrised product
\begin{equation*}
     \frac12\, \left(z^{-1}(\beta+\gamma)^2 + (\beta+\gamma)^2 z^{-1}\right)= -\frac{\chi}{2}\partial_\chi^2 - \partial_\chi \partial_\eta - \frac{1}{2\chi}\partial_\eta^2 - 2(l+1)\partial_\chi - \frac{4l+3}{2\chi}\partial_\eta - \frac{1}{\chi}\left(2l^2 + 3l + \frac{17}{8}\right)\ .
\end{equation*}
The operator on the right hand side is obtained by substituting (\ref{Fourier}-\ref{Fourier3}) and changing variables to $(\chi,\eta)$. Clearly, it preserves the space of functions periodic in $\eta$. This conclusion remains for any choice of operator ordering. In a similar manner, any classical function invariant under \eqref{identifications-Poincare} gives rise to an operator which respects periodicity \eqref{eta}.

Written in coordinates $(\chi,\eta)$ the scalar product on $\mathcal{H}\otimes\mathcal{\bar H}$ reads (see \eqref{inner-produt-Fourier})
\begin{equation}
    \langle f_1,f_2\rangle = 2^{4l+2}\pi^2 \iint d\eta\, d\chi\ \chi^{4l+3}\ \overline{f_1(\chi,\eta)} f_2(\chi,\eta)\ . 
\end{equation}
Since the measure does not depend on $\eta$, this inner product is also well-defined after the quotient by \eqref{eta}. The resulting Hilbert space of square-integrable functions will be denoted by
\begin{equation}\label{reduced-Hilbert-space}
    \mathcal{H}_{red} = L^2(\mathbb{R}\times S^1,\chi^{4l+3} d\chi d\eta)\ .
\end{equation}
In the following we will continue to work with representations with $\, l=\bar l\,$: the case with $\,l\neq \bar l\, $ may be considered as well, but we shall not do so.

\subsection{Spectrum of the radial coordinate}

We now turn to the BTZ radial coordinate \eqref{radius-BTZ}: for the remainder of this section, we will consider the closely related function $B(r)$. The expression for $\,B$ in  Poincar\'e coordinates holds in each of the regions I, II, III, \cite{Banados:1992gq}: in regions I outside the outer horizon ($r> r_+$) $\,B$ is positive, and in the black hole interior ($r< r_+$), $B$ is negative. In addition, the relation $r^2\geq 0\,$ in terms of $\,B$ gives the condition
\begin{equation}\label{minimum-B}
    B\geq -\frac{r_+^2}{r_+^2-r_-^2}\, \ .
\end{equation}
To define $B$ as an operator, we need to make a choice of the operator ordering as $\beta$ and $\gamma$ do not commute with $z$. We will use the symmetrised products:
\begin{equation}\label{symmetrised}
    \frac{\beta+\gamma}{z} := \frac12 \left\{\beta+\gamma,z^{-1}\right\} = - E_+^{-1}\Big(H-\frac12\Big) , \qquad \frac{\beta-\gamma}{z} := \frac12 \left\{\beta-\gamma,z^{-1}\right\} = - \bar E_+^{-1}\Big(\bar H - \frac12\Big)\ .  
\end{equation}
This gives
\begin{equation}
    B =  E_+^{-1} \Big(H-\frac12\Big)\,\bar E_+^{-1} \Big(\bar H - \frac12\Big)\ .  \\[2pt]
\end{equation}
In particular,  $B$  factorises into chiral and anti-chiral pieces. As differential operators, these are 
\begin{equation}\label{generators-of-the-boundary}
  \frac{\beta+\gamma}{z} = - i\left(\partial_x + \frac{l+\frac12}{x}\right), \qquad \frac{\beta-\gamma}{z} =-  i\left(\partial_{\bar x} + \frac{\bar l+\frac12}{\bar x}\right)\ .
\end{equation}
We are however more interested in the expression for $B$ in $(\chi,\eta)$ variables. In these coordinates the operator $B$ becomes
\begin{equation}
    B = \frac14\left( -\partial_\chi^2 - \frac{4l+3}{\chi}\,\partial_\chi - \frac{(2l+1)^2}{\chi^2} + \frac{1}{\chi^2}\,\partial_\eta^2\right)\ .
\end{equation}
By construction, $B$ commutes with $\, H-\bar H = \partial_\eta\,$. Therefore we can restrict ourselves to solving for eigenvectors of $B$ among functions with a fixed $\eta-$Fourier mode,
\begin{equation}
    B f_{n,\lambda}(\chi,\eta) = \lambda^2 f_{n,\lambda}(\chi, \eta)\, , \qquad f_{n,\lambda}(\chi, \eta) = e^{\frac{2\pi i n }{\alpha}\, \eta}\, f_\lambda(\chi), \quad n\in \mathbb{Z}\ .
\end{equation}
It is useful to write the resulting eigenvalue equation for $f_\lambda(\chi)$ in the Schr\"odinger form. Introducing
\begin{equation}\label{Sch1}
    f_\lambda(\chi) = \chi^{-2l-3/2}\, h_\lambda(\chi)\,,         
\end{equation}
we obtain equation
\begin{equation}\label{eigenh}
    -\frac{d^2h_\lambda\,}{d\chi^2} - \left(c^2+\frac 14\right)\, \frac{1}{\chi^2}\, h_\lambda =4\lambda^2\, h_\lambda\,,
\end{equation}
with
\begin{equation}\label{c}
     c=c(n)=- \frac{2\pi  n}{\alpha} = \frac{n \ell\,}{r_-}\ .
\end{equation}
Equation \eqref{eigenh} is the eigenvalue equation for a particle moving on a line in the attractive $\chi^{-2}$ potential,
\begin{equation}\label{V}
    V(\chi) =- \left(c^2+\frac 14\right)\,\frac{1}{\chi^2}\ .
\end{equation}
Solutions to \eqref{eigenh} for positive eigenvalues $\lambda^2$ and fixed $ c$, the  scattering states of the potential \eqref{V}, are the Bessel functions of imaginary order which we will write in terms of the Hankel functions  $H^{(1,2)}_{ic} $,
\begin{equation}
    f_{\lambda}(\chi) = \sqrt{\chi} \left(C_1 \, H^{(1)}_{ic}(2\lambda\chi) + C_2\, H^{(2)}_{ic}(2\lambda\chi)\right)\ .
\end{equation}
They behave asymptotically  as  plane waves,
\begin{equation}
   \sqrt{\chi}\,  H_{ic}^{(1,2)}(2\lambda\chi)\sim  \frac{1\pm i}{\sqrt{2\pi\lambda}}\ e^{\pm\left(2i\lambda\chi+\frac{c\pi}{2}\right)}\,, \quad \chi\to\infty\,,
\end{equation}
and vanish at $\chi= 0$. There are continuously many scattering states, that is, continuously many eigenstates of $r^2$  outside the outer horizon, with eigenvalues $\ r_+^2+\lambda^2(r_+^2 - r_-^2)\,$.

States of the BTZ black hole inside the horizon $ r_+$ are the eigenstates of $B$ with $\lambda^2<0$, i.e. the bound states of the potential \eqref{V}. With regard to the bound states, the attractive $\chi^{-2}$ potential is usually considered  unphysical. The reason can be  seen already in equation \eqref{eigenh}: equal scaling of the kinetic and the potential terms implies that, if $\, h_\lambda(\chi)$ is a solution, then so is the whole family $\, h_{\lambda/\mu} (\mu\chi)$, $\mu\in\mathbb{R}_+$. Applied to bound states this means that if there is one bound state, there is a continuum of them. Given that function $H^{(1)}_{ic}$  exponentially decreases at infinity i.e. that it is normalisable
\begin{equation}
   \sqrt{\chi} \, H^{(1)}_{ic}(2i\lambda\chi) =  \sqrt{\chi}\, K_{ic}(2\kappa\chi)\sim \frac12\sqrt{\frac{\pi}{\kappa}}\  e^{-2\kappa\chi} \,, \quad  \chi\to\infty\,,
\end{equation}
($ \kappa^2=-\lambda^2$, $\,\kappa>0$)\,, the bound states exist. Mathematically, the problem arises because $B$ is not a self-adjoint operator, but only formally self-adjoint, \cite{Hutson}.\footnote{The chiral and anti-chiral parts of $B$ are also formally self-adjoint, but in  intervals $\,x\in [0,\infty)$, $\,\bar x\in [0,\infty)\,$ they do not have self-adjoint extensions. This fact shows up in non-orthogonality of their eigenfunctions: we have, for example
\begin{equation*}
    (x^{-l-1/2} e^{i\kappa_1 x}, x^{-l-1/2} \, e^{i\kappa_2 x}) = 2^{2l+1}\pi \int\limits_0^\infty dx\ e^{i(\kappa_2-\kappa_1) x} = \frac{2^{2l+1}\pi i}{\kappa_2 - \kappa_1}\,,
    \quad \text{for}\quad  \kappa_1\neq\kappa_2\ .       
\end{equation*}}
Self-adjoint extensions of $B$ can be obtained by imposing the appropriately chosen boundary conditions, \cite{x^-2}. This, however, does not resolve the problem, as each choice of a self-adjoint extension gives a different spectrum of $B$. In \cite{Landau}, this property of the bound state spectrum is interpreted as  feature of the attractive inverse-square potential that all negative-energy particles fall to the centre ($\chi =0$, $\, B=-\infty\,$).

In our analysis, $B$ is related to the radius $r$ via \eqref{B(r)}. But $r$ is the radial (real) coordinate, and we should ensure that $r^2\geq 0\,$ holds at the quantum level. In consequence, the eigenvalues of $B$ are to be constrained as
\begin{equation}
-\kappa^2=    \lambda^2\geq -\, \frac{r_+^2}{r_+^2-r_-^2}\ ,
\end{equation}

\begin{figure}[ht]
    \centering
        \begin{minipage}{0.45\textwidth}
        \centering
        \includegraphics[scale=0.5]{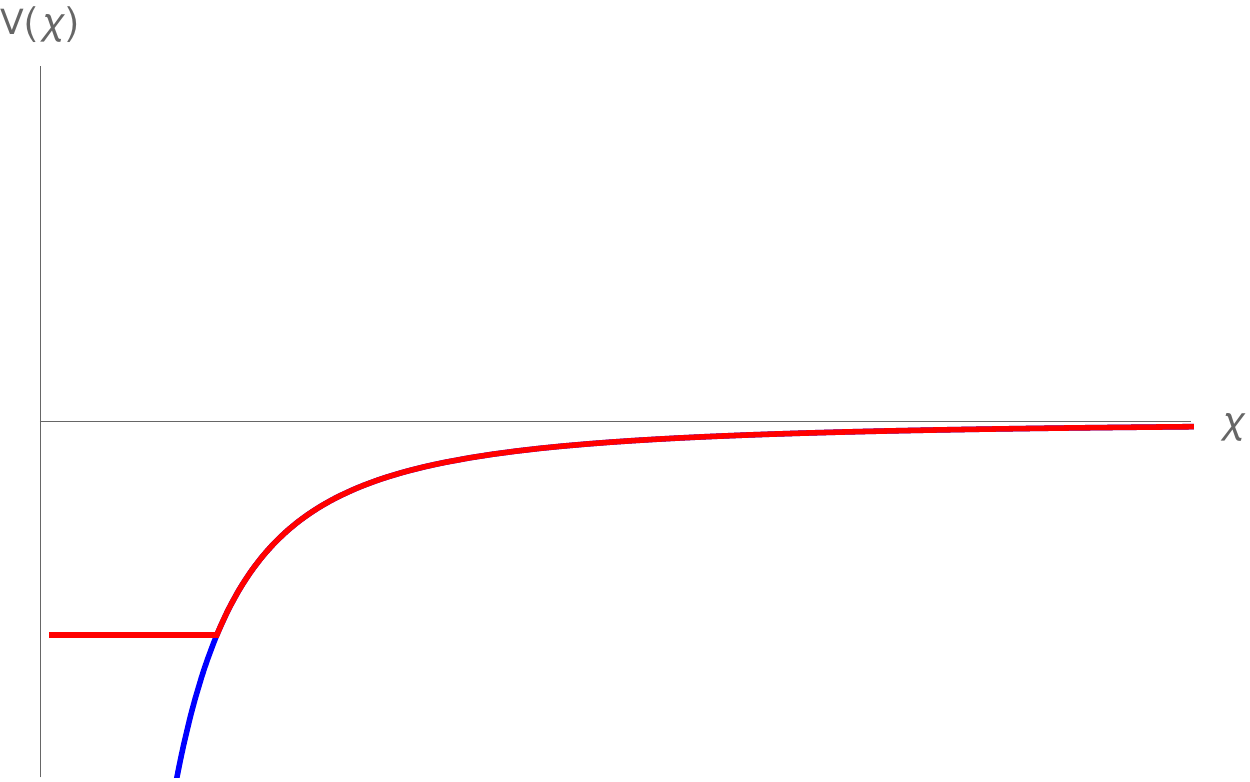}
    \end{minipage}
    \begin{minipage}{0.45\textwidth}
         \centering
        \includegraphics[scale=0.5]{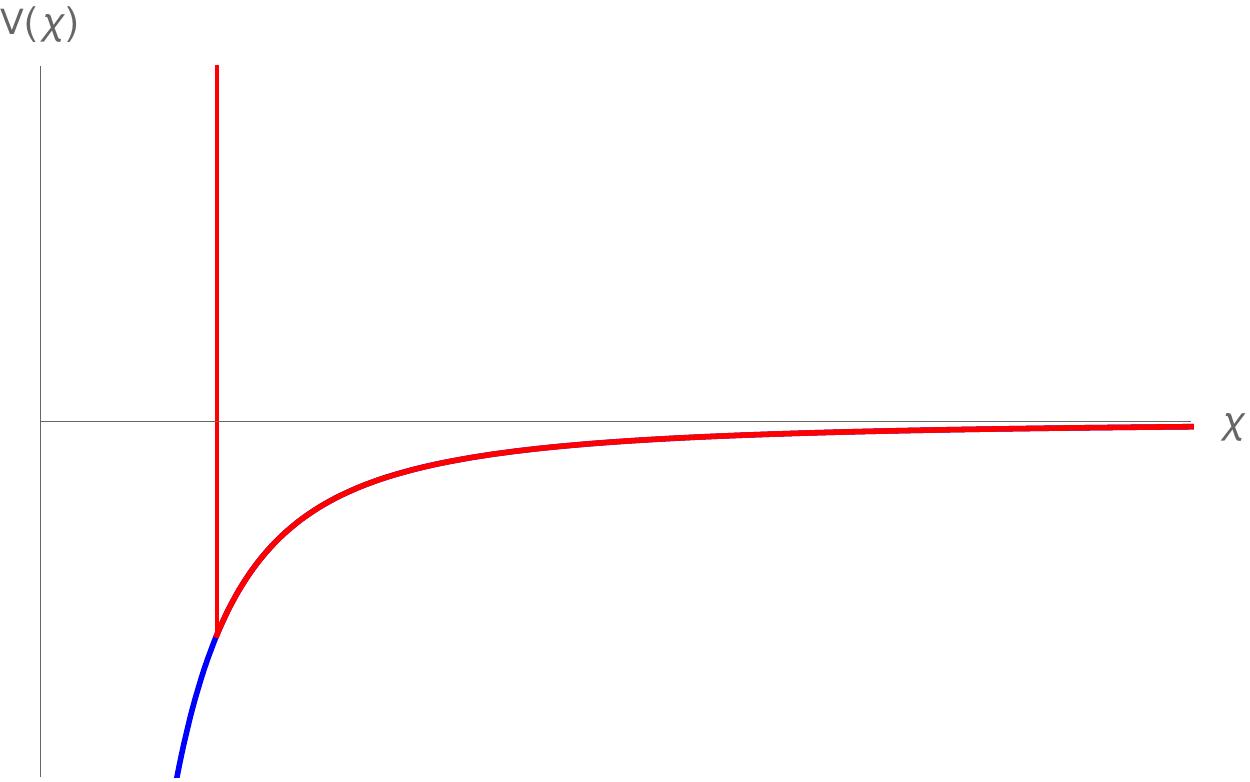}
    \end{minipage}
    \caption{Inverse square potential and its modifications}
\end{figure}
that is they have a (negative) lower bound. A natural way to enforce this condition is to modify the potential $\,V(\chi)$ by giving it a minimal value, $\, V_{min} = -4\kappa_0^2\,$. The simplest modifications considered in the literature \cite{Landau,x^-2}, given in Figure~1, are
\begin{equation}\label{Vmod}
    V^\star(\chi) =\left\{ \begin{array}{ll}
      -\, \dfrac{4 r_+^2}{r_+^2-r_-^2} \, , \quad  &  \chi\in (0, \chi_0)  \\[12pt]
     - \left(c^2+\dfrac 14\right)\,\dfrac{1 }{\chi^2} \,, \ \    &  \chi\geq\chi_0 
    \end{array}   \right. 
    \ \ \text{and} \quad 
     V^\ast(\chi) =\left\{ \begin{array}{ll}
      \infty \, , \quad  &  \chi\in (0, \chi_0)  \\[12pt]
     - \left(c^2+\dfrac 14\right)\,\dfrac{1 }{\chi^2} \, , \ \    &  \chi\geq\chi_0 
    \end{array}   \right.   
\end{equation}
where in both cases
\begin{equation}
   4\kappa_0^2=  \frac{4 r_+^2}{r_+^2-r_-^2}  = \left(c^2+\frac 14\right)\, \frac{1}{\chi_0^2}\ .\\[2pt]
\end{equation}
We discuss the eigenstates corresponding to these potentials in Appendix C. Both modifications have the same qualitative properties: for a fixed $c$ there is a discrete infinite set of bound states whose eigenvalues exponentially accumulate at $\lambda^2\to 0-\, $. This means that fuzzy BTZ black hole has an infinite discrete set of states inside the outer horizon, $r_+\,$.
 
\section{Semi-classical states}

A common way to relate a noncommutative space to the corresponding classical manifold is to identify a set of coherent states in the Hilbert space, which behave in a way similar to classical points and provide a notion of local measurements. In this section we will propose such a set of semi-classical states on the fuzzy AdS$_3$. The method shares similarities with generalised coherent states of \cite{Perelomov:1986tf}, but differs from it in several points. In the limit when the quantum numbers of discrete series $l=\bar l\to-\infty$ the relative uncertainties of coordinates in these states tend to zero, thereby showing that this may be regarded as the classical limit.

In order to define semi-classical states, we start from the lowest weight vector in the Hilbert space $\mathcal{H} \otimes \mathcal{\bar H}$, denoted $|0\rangle$. This is the tensor product of lowest weight vectors \eqref{lowest-weight} of the two factor representations of $SL(2,\mathbb{R})$. In our realisation it is given by
\begin{equation}
    |0\rangle = C^2 (x\bar x)^{-2l-1} e^{-x-\bar x}\, ,\qquad C^{-2} =  2^{4l+1}\pi \,\Gamma(-2l)\ . 
\end{equation}
A set states $|\xi\rangle$ labelled by points of the classical AdS$_3$ space will be obtained by acting on $|0\rangle$ with transformations generated by momenta. Note that expectation values of Poincar\'e coordinates in the lowest weight state are
\begin{equation}
    z_0 = \langle0|z|0\rangle =  \frac{\Gamma (1-a-2l) \Gamma (a-2l)}{\Gamma(-2l)^2}, \quad \beta_0 = \langle0|\beta|0\rangle = 0, \quad \gamma_0 = \langle0|\gamma|0\rangle = 0\ .
\end{equation}

Let $\,\xi_0$ be the classical point of AdS$_3$ with coordinates $(z_0,\beta_0,\gamma_0)$. At least locally, any other point may be obtained from $\xi_0$ by a transformation of the form
\begin{equation}\label{action-on-classical-points}
    \xi = \lambda^{-\tilde p_z} e^{-b \tilde p_\beta} e^{-c \tilde p_\gamma}\, \xi_0 = (\lambda^{-1}z_0,\beta_0 - b z_0,\gamma_0 - c z_0)\,,
\end{equation}
where, recall, $\{\tilde p_z,\tilde p_\beta,\tilde p_\gamma\}$ is the classical moving frame
\begin{equation}
    \tilde p_z = \tilde z \partial_{\tilde z}, \quad \tilde p_\beta = \tilde z \partial_{\tilde \beta},\quad \tilde p_\gamma = \tilde z \partial_{\tilde \gamma}\ .
\end{equation}
We define the state $|\xi\rangle$ corresponding to the above point by
\begin{equation}
   |\xi\rangle = |\lambda,b,c\rangle  = \lambda^{p_z} e^{b p_\beta} e^{c p_\gamma} |0\rangle\ .
\end{equation}
The first observation about the state $|\xi\rangle$ is that expectation values of Poincar\'e coordinate operators in it coincide with the classical values of these coordinates. Indeed,
\begin{equation*}
    \langle \xi| z |\xi\rangle = \langle 0| e^{-c p_\gamma} e^{-b p_\beta} \lambda^{-p_z} z \lambda^{p_z} e^{b p_\beta} e^{c p_\gamma}| 0\rangle = \langle 0| e^{-c p_\gamma} e^{-b p_\beta} \lambda^{-1} z e^{b p_\beta} e^{c p_\gamma}| 0\rangle = \lambda^{-1} \langle 0| z |0\rangle\ .
\end{equation*}
The above calculation follows from the BCH formula once the frame relations \eqref{momenta-coords} are taken into account. The same argument applies to other Poincar\'e coordinates, leading to
\begin{equation}\label{semi}
    \langle \xi| z |\xi\rangle = \lambda^{-1}z_0, \quad \langle\xi| \beta |\xi\rangle = \beta_0 - b z_0, \quad \langle\xi| \gamma |\xi\rangle = \gamma_0 - c z_0\,,
\end{equation}
in accord with \eqref{action-on-classical-points}. Note that the last argument is independent of the choice of the initial state $|0\rangle$ and would apply to other systems of states generated in the same way from a different vector $|0'\rangle$. For the boundary coordinates \eqref{symmetrised}, there is also an exact agreement between expectation and classical values
\begin{equation}
    \langle\xi|\,\frac{\beta+\gamma}{z}\, |\xi\rangle = -\lambda(b+c), \quad \langle\xi| \,\frac{\beta-\gamma}{z}|\,\xi\rangle = -\lambda(b-c)\ .
\end{equation}

Clearly, arbitrary functions of $z$, $\beta$, $\gamma$ cannot satisfy such relations exactly. The difference between the quantum and the classical system can be quantified by relative uncertainties of operators. For uncertainties of the Poincar\'e coordinates we find
\begin{align}
    & \frac{\langle\xi|z^2|\xi\rangle - \langle\xi|z|\xi\rangle^2}{\langle\xi|z^2|\xi\rangle} = 1-\frac{\Gamma (1-a-2l)^2 \Gamma (a-2 l)^2}{\Gamma (-2 l)^2 \Gamma (2 (a-l)) \Gamma (-2 (a+l-1))} \to 0 \quad \text{as} \ \ l\to-\infty\,,\\[6pt]
    & \frac{\langle\xi|\beta^2|\xi\rangle -\langle\xi|\beta|\xi\rangle^2 }{\langle\xi|\beta^2|\xi\rangle} = 1-\frac{8 b^2 (2 l+1) \Gamma \left(\frac{1}{2}-2 l\right)^4}{\left(4 l \left(4 l \left(b^2 (4 l+2)-1\right)-1\right)+1\right) \Gamma (-2 l)^4} \to 0 \quad \text{as} \ \ l\to-\infty\,,\ b\neq0\,,\\[6pt]
    & \frac{\langle\xi|\gamma^2|\xi\rangle -\langle\xi|\gamma|\xi\rangle^2 }{\langle\xi|\gamma^2|\xi\rangle} = 1-\frac{8 c^2 (2l+1) \Gamma \left(\frac{1}{2}-2 l\right)^4}{\left(8 l (2 l+1) \left(4 c^2 l-1\right)-1\right) \Gamma (-2 l)^4}\to 0 \quad \text{as} \ \ l\to-\infty\,,\ c\neq0\ .
\end{align}
We wrote, for the sake of simplicity, the relative uncertainties of $\beta$ and $\gamma$ for the AdS$_3$ value $a=1/2$, \eqref{ads}; the expressions for arbitrary $a$ also vanish in the $l\to-\infty$ limit. The same holds for functions of Poincar\'e coordinates, e.g.
\begin{equation}
     \frac{\langle\xi|z^{-2}|\xi\rangle - \langle\xi|z^{-1}|\xi\rangle^2}{\langle\xi|z^{-2}|\xi\rangle} = 1-\frac{\Gamma (-a-2 l)^2 \Gamma (a-2 l-1)^2}{\Gamma (-2 l)^2 \Gamma (2 a-2 l-2) \Gamma (-2 (a+l))}\to 0, \quad \text{as} \ \ l\to-\infty\ .
\end{equation}
Therefore we see that the semi-classical state $|\xi\rangle $ approximates the classical point $\xi$ with improving accuracy as $l\to-\infty$, which, taken together with \eqref{semi}, means that $l\to-\infty\,$ gives the classical limit of the fuzzy AdS$_3$ space.

Semi-classical states are not mutually orthogonal. Their overlap functions are 
\begin{equation}\label{overlaps}
    \langle \xi_1|\xi_2\rangle = 16^{-l}\left(\frac{(\lambda_1 c_1-\lambda_2 c_2)^2-(\lambda_1 b_1-\lambda_2 b_2)^2+(\lambda_1+\lambda_2)^2}{\lambda_1\lambda_2} - 2i \frac{(\lambda_1 + \lambda_2)(\lambda_1 b_1 - \lambda_2 b_2)}{\lambda_1\lambda_2}\right)^{2l}\ .
\end{equation}
Further, it can be shown that, unlike standard coherent states, $|\xi\rangle$ do not provide a resolution of unity, a property shared with the set of `weak coherent states' discussed in \cite{Klauder}. Since we do not use the states $|\xi\rangle$ to define the noncommutative product as done e.g. in \cite{Perelomov:1986tf,Grosse:1993uq}, the resolution of unity has no immediate interpretation and is not required for consistency.\footnote{If one uses coherent states to define operators corresponding to classical functions in the  usual way, resolution of unity is the statement that constant functions are mapped to multiples of the identity operator.} Nevertheless, a relevant question which we leave for future research, is whether the kernel \eqref{overlaps} is algebraically special in some sense. It is interesting to observe that the real part of the expression inside the bracket is (a function of) the AdS$_3$ geodesic distance between points $\Phi(\xi'_1)$ and $\Phi(\xi'_2)$, where $\xi_i = \lambda^{-\tilde p_z} e^{-b \tilde p_\beta} e^{-c \tilde p_\gamma}(i,0,0)$,
\begin{equation}
    \text{Re}\left(\langle\xi_1|\xi_2\rangle^\frac{1}{2l}\right) = \cosh^2 \left(\frac{d(\Phi(\xi'_1),\Phi(\xi'_2))}{2}\right)\,,
\end{equation}
where, $\Phi$ was defined in \eqref{beta'}.

\section{Summary and outlook}

In this work, we have constructed noncommutative models of the three-dimensional anti-de~Sitter space and the BTZ black hole. We shall shortly summarise the obtained results and discuss some future research directions. 

Our starting point was the set of expressions (\ref{Z1}-\ref{G1}) that define noncommutative analogues of the Poincar\'e coordinates, together with a moving frame, as operators on a Hilbert space. While such a definition is in part a matter of choice, we showed that it allows to develop differential geometry closely analogous to the commutative one. The noncommutative metric \eqref{metr}, essentially fixed by the frame, admits a compatible torsion-free connection \eqref{connection}, and the associated curvature satisfies vacuum Einstein's equations with a negative cosmological constant. We derived the Laplace-Beltrami operator, including its action on arbitrary differential forms, (\ref{laplacian-functions}-\ref{lap-3forms}). The boundary of the fuzzy AdS$_3$ is found to be commutative and flat, \eqref{boundary}.

Noncommutative coordinates (\ref{Z1}-\ref{G1}) allow for a free parameter $a$, which does not enter into structure functions of the differential calculus. We showed that $a$ may be related to the global spacetime structure: requiring that the effect of BTZ identifications on Poincar\'e coordinates is implemented on the quantum level by a unitary operator $\,U=e^{\alpha(H-\bar H)}\,$ fixes $a$ and $\alpha$ uniquely, \eqref{aalpha}. Relations \eqref{aalpha} show how the model depends on the parameters that characterise the classical BTZ black hole, the radii of its inner and outer horizons. Notice however that $a$ and $\alpha$ are not defined for $\, r_-=0$, that is, the given construction does not describe the non-rotating fuzzy black hole. This is, in a way, in accord with the fact that non-rotating and rotating BTZ black holes have different classical geometries, \cite{Banados:1992gq}. On a related note,  extending $r_\pm$ beyond physical values and putting $r_+=0$, $\, r_-=i\ell$, we formally obtain
\begin{equation}
    M=-1\, , \quad a=\frac 12\, ,\quad \alpha =-2\pi i\ ,                  \label{ads}
\end{equation}
which makes the BTZ identifications \eqref{U} trivial. In this sense, the value $a=1/2\,$  describes the fuzzy AdS$_3$ space, with $\, B(r) =r^2/\ell^2$. 
 
Among the most important properties of a noncommutative space are the spectra of physically relevant coordinates. We analysed here the radial coordinate $r$. In computations, we used the Fourier space realisation of discrete series representations, (\ref{Fourier}-\ref{Fourier3}): in this realisation, identifications by $U$ are implemented by making one coordinate periodic, \eqref{eta}.
The eigenvalue equation for $B(r)$, \eqref{B(r)} can be written in a Schr\"odinger form \eqref{eigenh}, upon which eigenstates above and inside the horizon at $r_+$ are identified with the scattering and bound states. The spectrum of scattering states is continuous and that of bound states is infinite and discrete. It might be worth noting that the fact that the outer horizon $r=r_+$ separates the continuous and discrete parts of the spectrum comes out of a computation, rather than being satisfied by construction. Bound states accumulate at $\, r_+\vert_{-0}\,$ exponentially, \eqref{asymptotic-energies}. The treatment of bound states required regularisation, that could be seen physically as excluding values $\,r^2<0\,$ by a modification of the potential, \eqref{Vmod}: we may regard the regularisation procedure as a part of the quantisation prescription. Luckily, by all accounts, different procedures, of which we studied two, lead to same qualitative results.

In the last section we introduced a set of semi-classical states in the Hilbert space labelled by points of the classical AdS$_3$ manifold. These states are constructed in such a way that expectation values of Poincar\'e coordinate operators in them equal the values of corresponding classical functions. It was shown that relative uncertainties of coordinates in semi-classical states vanish for $l\to-\infty$, thus providing the classical limit of our fuzzy space. 

One feature of the above model that should be mentioned is that there is no isomorphism between the commutative algebra of functions $\mathcal{\tilde{A}}$ and its noncommutative replacement $\mathcal{A}$. This property is for example satisfied by the fuzzy sphere in a certain limit. The question may be analysed by decomposing the noncommutative operator algebra into irreducible representations of the symmetry group and comparing the result with the geometric representation on the space of functions of the classical spacetime. Let us discuss this question for the AdS$_3$, which has the unbroken $SO(2,2)$ symmetry. Due to the isometry AdS$_3\cong SL(2,\mathbb{R})$, the space of functions on AdS$_3$ decomposes as the regular representation\footnote{By $\pi^\ast$ we denote the dual, i.e. the contragredient representation of $\pi$.}
\begin{equation}\label{Peter-Weyl}
    \mathcal{\tilde A} = L^1(SL(2,\mathbb{R})) = \sum_{\pi\in PD} \text{End} (\pi) = \sum_{\pi\in PD} \pi\otimes\pi^\ast\,,
\end{equation}
where $PD$ is the set of principal and discrete series unitary irreducible representations of $SL(2,\mathbb{R})$, described in Appendix B. More precisely, matrix elements $\pi_{ij}\,$ span a dense subspace of $L^1(SL(2,\mathbb{R}))$. Notice that we have written the decomposition into left-right bimodules of $SL(2,\mathbb{R})$, or equivalently, representations of $SL(2,\mathbb{R})\times SL(2,\mathbb{R})$. On the other hand, the algebra $\mathcal{A}$ can be decomposed into irreducibles of $SL(2,\mathbb{R})$ using results of \cite{Repka} that we collect in Appendix B. The decomposition reads
\begin{equation}\label{quantum-space-deocomposition}
    \mathcal{A} = \text{End}(\mathcal{H} \otimes \mathcal{\bar H}) = (\mathcal{H}\otimes\mathcal{H}^\ast)\otimes(\mathcal{\bar H}\otimes\mathcal{\bar H}^\ast) = \int_{\mathbb{R}_+^2} d\rho d\bar\rho\ T_{i\rho-1/2,0}\otimes T_{i\bar\rho-1/2,0} = L^2(\mathbb{H}^2 \times \mathbb{H}^2)\ .
\end{equation}
In the last line, it is understood that hyperbolic spaces $\,\mathbb{H}^2$ carry the standard measure $y^{-2}dx\, dy$. We see that the classical and quantum spaces do not coincide, the quantum being of `one dimension higher'.

It is of interest to determine if the additional dimension of the quantum space is compact, in which case it would give rise to a discrete set of Kaluza-Klein modes. Note that it was argued in \cite{Sperling:2019xar,Steinacker:2019fcb} that Kaluza-Klein modes are a generic feature of quantum spaces in more than two dimensions and these were interpreted in terms of higher spin fields. We are really interested in the BTZ case, where indeed the additional dimension seems to be compact for the following reason. Since in the BTZ geometry $SO(2,2)$ symmetry is broken, one cannot compare classical and quantum spaces in terms of their $SO(2,2)$ irreducible content. Instead, we note that after the quotient by identifications \eqref{eta}, the Hilbert space $\mathcal{H}_{red}$ of the quantum system is that of functions one cylinder, \eqref{reduced-Hilbert-space}. Operators $\mathcal{O}$ on $\mathcal{H}_{red}$ may be encoded by integral kernels $K(\chi,\eta,\chi',\eta')$ such that their action on functions reads
\begin{equation}
    \mathcal{O}[f](\chi,\eta) = 2^{4l+2}\pi^2  \int d\eta' d\chi'\ \chi'^{4l+3}\, K(\chi,\eta,\chi',\eta') f(\chi',\eta')\ .
\end{equation}
The kernels belong to the appropriate space of functions $C(\mathbb{R}\times S^1\times \mathbb{R}\times S^1)$ on the trivial $S^1$ fibre bundle over the BTZ topology (the topology of the BTZ black hole is $\mathbb{R}^2\times S^1$). This suggests that the quantum space indeed has one compact dimension more than the classical one and quantum operators correspond to functions on the black hole together with an infinite discrete set of $S^1$ Kaluza-Klein modes. To uncover this structure, the first step would be to identify inside $\mathcal{A}_{red}=\text{End}(\mathcal{H}_{red})$ the quantisations of classical functions. For this, a general ordering prescription needs to be chosen. After identifying quantised functions, one would try to organise the remaining operators in $\mathcal{A}_{red}$ into KK modes. We will investigate this question further in future work.

The last question will also inevitably be addressed when constructing quantum field theories over $\mathcal{A}_{red}$. A necessary prerequisite for investigations of field theories is the knowledge of the Laplacian and its eigenfunctions. We have defined this operator in Section 3. The Laplacian acts on the algebra of functions $\mathcal{A}$ and more generally that of differential forms $\Omega^\ast(\mathcal{A})$, but it also acts on the Hilbert space of the theory itself. In the coordinates $(\chi,\eta)$, acting on functions \eqref{Sch1}, the operator is independent of $\eta$ (thus being well-defined on $\mathcal{H}_{red}$) and assumes the very simple form
\begin{equation}
    \Delta = -\chi^2 \partial_\chi^2 + \left(4\chi^2 + \frac34\right) \sim -\partial_X^2 + (4e^{2X} + 1)\ .
\end{equation}
By the last line, we mean that one may bring $\Delta$ to the rightmost form by changing the variable as $\chi = e^{X}$ and performing a similarity transformation. The eigenfunctions of $\Delta$ are readily constructed in terms of Bessel functions. We will solve the more difficult eigenvalue problem for $\Delta$ on $\mathcal{A}_{red}$ in the future work. The resulting fuzzy harmonics will serve as the starting point for developing quantum field theory on $\mathcal{A}_{red}$. Besides quantum field theory on the fixed fuzzy background $\mathcal{A}_{red}$, an obvious challenge is to embed $\mathcal{A}_{red}$ as a ground state of a dynamical theory of gravity. Probably the right framework to attempt this are matrix models, similarly to what was done in \cite{Jurman:2013ota,Sperling:2019xar,Steinacker:2016vgf}.

Several other problems can and should be addressed in the given framework. One is to understand properties of the inner horizon $r_-\,$: in our model, the point $r=r_-$ is in no way special and likely to be absent from the spectrum of the radial coordinate. A related issue is to establish a model of the fuzzy non-rotating BTZ black hole, $r_-=0$, either by using another set of classical coordinates, or by a different quantisation of $\,(z,\gamma,\beta)$. 

Next, we would like to further investigate relations of semi-classical states to classical geometry, as might be encoded in their overlaps and matrix elements of geometric operators. On a related note, one should study the role of the initial vector $|0'\rangle$ on properties of associated semi-classical states, as well as the extension of the construction to the BTZ fuzzy geometry.

Finally, a problem that deserves a separate study is the entropy of the fuzzy BTZ black hole. In the exact quantum gravity solution we would expect that the number of black hole states below the horizon is finite, providing the known value of the black hole entropy. In our model all below-horizon states are discrete, but their number is infinite. The endeavour to refine or modify the present model by a further regularisation or restriction of parameters $l$, $\bar l\,$, $r_\pm$, $\ell\,$ in relation to the black hole entropy is a very important further task.

\vskip10pt
{\bf Acknowledgements:} This work is  funded by a research grant under the project H2020 ERC STG 2017 G.A. 758903 "CFT-MAP"  and by 451-03-9/2021-14/200162 Development Grant of MPNTR, Serbia. We wish to thank the anonymous referee for useful remarks and suggestions.

\appendix 

\section{Coordinate systems}

In this appendix we collect various formulas about different coordinate systems for AdS$_3$ and other spaces locally isometric to it. We mainly follow \cite{Banados:1992gq,Carlip:1995qv}.

\subsection{Anti-de Sitter space and its universal cover}

The AdS$_3$ space is the hyperboloid $\  -v^2-u^2+x^2+y^2= -\ell^2    \ $ in the flat four-dimensional space
\begin{equation}
 ds^2=-dv^2-du^2+dx^2+dy^2  \, , \quad v,\,   u,\,  x, \,  y\in(-\infty,\infty)\  .
\end{equation}
A global, static coordinate system $(\tau,\rho,\theta)$ on the hyperboloid, up to trivial singularities of angular coordinates, may be introduced by
\begin{equation} \label{tau}
 v=\ell\cosh \rho \cos \tau, \quad u=\ell\cosh \rho \sin \tau, \quad x=\ell \sinh \rho \cos \theta, \quad y=\ell \sinh \rho \sin \theta\ .
\end{equation}
 The line element  reads
\begin{equation}
 ds^2 = \ell^2(-\cosh^2\rho\, d\tau^2 + d\rho^2 + \sinh^2\rho\, d\theta^2) \, ,\quad \rho\in[0,\infty),  \  \theta,\tau\in[0,2\pi)\ .
\end{equation}
The time $\tau$ as defined in \eqref{tau} is  periodic, and one usually unwraps the closed timelike curves by not identifying $\,\tau$ with $\,\tau + 2\pi n$, $\,n\in \mathbb{Z}$. By the common abuse of terminology, we refer to the resulting universal cover as anti de-Sitter space as well, and denote it by $\widetilde{\text{AdS}_3}$. 

The line element  in the Schwarzschild form is obtained if we introduce $\,   r=\ell \sinh \rho$, $\, t = \ell\tau\, $. Then
\begin{equation}\label{8}
 ds^2 =- \left(\frac{r^2}{\ell^2}+1\right) dt^2 +\frac{1}{\ \dfrac{r^2}{\ell^2}+1\  }\, dr^2 +r^2 d\theta^2\,, \quad r\in[0,\infty),\   t\in( -\infty ,\infty), \ \theta\in[0,2\pi)\ .    
\end{equation}
Both AdS$_3$ and $\,\widetilde{\text{AdS}_3}$ can be conformally compactified at $\,r\to\infty$. Reparametrising the radial coordinate according to $\,\cosh\rho = (1+\sigma^2)/(1-\sigma^2)$, $\, \sigma = \tanh (\rho/2)$, we obtain
\begin{equation}                              \label{A5}
    ds^2 = \frac{\ell^2}{(1-\sigma^2)^2}\, \Big(-(1+\sigma^2)^2 d\tau^2 + 4d\sigma^2 + 4\sigma^2 d\theta^2\Big) \equiv \frac{\ell^2}{(1-\sigma^2)^2} \, d\tilde s^2
    \, ,\quad \sigma\in[0,1),\  \theta,\tau\in[0,2\pi)\ .
\end{equation}
The metric diverges as $\sigma\to 1$, where the conformally equivalent metric $d \tilde s^2\,$ is flat. The set $\,\{\sigma=1\}$ is the conformal boundary.

The Poincar\'e coordinates $(\gamma,\beta,z)$  are introduced as \eqref{Poincare}. The image of the AdS$_3$ hyperboloid under the map $\, (v,u,x,y) \mapsto(\gamma,\beta,z)\,$ consists of all  points with $z\neq 0$. If we restrict to one of the two half-spaces, e.g. $z>0$,  the coordinates cover a half of the hyperboloid. The universal covering space $\widetilde{\text{AdS}_3}$ requires an infinite number of Poincar\'e patches. In  Poincar\'e  coordinates the AdS$_3$ metric assumes conformally flat form,
\begin{equation}
    ds^2 = \frac{\ell^2}{z^2}\, ( -d\gamma^2 + d\beta^2 + dz^2)\,, \quad z\in [0,\infty)\,, \ \beta\,,\gamma\in(-\infty,\infty)\ .
\end{equation}
The  inverse  of \eqref{Poincare} is
\begin{equation}\label{linePoincare} 
    u=\dfrac{\ell}{2z}\,(1+z^2+\beta^2-\gamma^2), \quad x= \dfrac{\ell}{2z}\,(1-z^2-\beta^2+\gamma^2), \quad v=-\dfrac \ell z \, \gamma \, ,\quad y=\dfrac \ell z\, \beta\,,       
\end{equation}
  while for the radial polar coordinate  $r^2\,$ in AdS$_3$ we find
\begin{equation}\label{radius-AdS}
    r^2 = \dfrac{\ell^2}{4z^2}\,\Big( (1-z^2-\beta^2+\gamma^2)^2 + 4\beta^2\Big)\ .  
\end{equation}

\subsection{BTZ black hole and Poincar\'e patches}

The BTZ space is a rotating black hole of mass $M$ and angular momentum $J$. Its line element, written in polar coordinates, is given in \eqref{BTZ-metric}. The black hole has two horizons, outer $ r_+$ and inner $r_-\,$,  given by \eqref{horizons}. For the extreme black hole ($J=M\ell$) the two horizons merge, $r_+=r_-\, $. For the non-rotating black hole, $J=0$, $r_- = 0$, the line element is
\begin{equation}
    ds^2=-\left(\frac{r^2}{\ell^2}-M \right)\, dt^2 + \frac{1}{ \ \dfrac{r^2}{\ell^2}-M \ }\, dr^2 +r^2 d\phi^2\ .  \label{BTZ1}
\end{equation}
The vacuum of the family of BTZ black holes that is parametrised by real parameters $(M,J)$ is given by $M=0$, $J=0$. By replacing formally  $\, M=-1$, $J=0\,$,   we obtain the AdS$_3$ space \eqref{AdS}, a point disconnected from the family of physical BTZ spaces.

The BTZ black hole is locally isometric to $\,\widetilde{\text{AdS}_3}$ and can be obtained via a discrete quotient. It  arises  by identifying points $X$ under the action of the discrete subgroup of isometries $\,\Gamma\cong\mathbb{Z}$,
\begin{equation}
     X\to e^{2\pi n\,\xi}\,X ,\qquad n\in \mathbb{Z}\,,
\end{equation}
where $\xi$ is a Killing vector field. In the non-extreme case, $r_+\neq r_-\,$,  $\xi\,$ is equivalent to $\xi'$, \cite{Banados:1992gq}
\begin{equation}
   \xi' = \frac{r_+}{\ell}\, M_{12} -\frac{r_-}{\ell}\, M_{03}\ .
\end{equation}
The BTZ black hole is the quotient of the subset $\,\widetilde{\text{AdS}'_3}\subset \widetilde{\text{AdS}}_3$ on which $\xi'$ is spacelike, $\xi'\cdot \xi'>0$. Cutting out the regions $\,\xi'\cdot\xi'<0$ from $\,\widetilde{\text{AdS}_3}$ ensures that there are no closed timelike curves in the quotient; it is also responsible for geodesic incompleteness of the BTZ space $\,\mathcal{M} =\widetilde{\text{AdS}'_3} /\Gamma$. From the quotient perspective,  parameters $\,r_\pm$ correspond to the choice of the vector field $\,\xi'$.

Polar coordinates of the BTZ space are not those defined by projection from the corresponding coordinates in AdS$_3$. Therefore, relations between  polar and  Poincar\' e coordinates in the BTZ geometry are different from those in AdS$_3$. They are conveniently written in terms of variables
\begin{equation}
    \tilde{t} = \frac{r_+}{\ell^2} \,t- \frac{ r_-}{\ell}\,\phi\ , \qquad \tilde{\phi} = -\frac{r_-}{\ell^2}\,  t+ \frac{ r_+}{\ell}\,\phi \  ,\\[2pt]
\end{equation}
and depend on the region of the BTZ space one is considering. In any  patch there are three regions,
\begin{align*}                      \label{BTZemb}
    & \text{I}: && r>r_+ \quad  && u=\sqrt{A}\,\cosh\tilde\phi ,\ \   && x=\sqrt{A}\,\sinh\tilde\phi, \ \ && y=\sqrt{B}\,\cosh\tilde t  , && v=\sqrt{B}\,\sinh\tilde t\,, \\[4pt]
    & \text{II}: && r_+>r>r_- \quad  && u=\sqrt{A}\,\cosh\tilde\phi , && x=\sqrt{A}\,\sinh\tilde\phi , && y=-\sqrt{-B}\,\sinh\tilde t  , && v=-\sqrt{-B}\,\cosh\tilde t\,,     \\[4pt]
    & \text{III}: && r_->r>0 && u=\sqrt{-A}\,\sinh\tilde\phi , && x=\sqrt{-A}\,\cosh\tilde\phi , && y=-\sqrt{-B}\,\sinh\tilde t  , && v=-\sqrt{-B}\,\cosh\tilde t\,,
\end{align*}
with $A=A(r)$ and $B=B(r)$ defined as
\begin{equation}
     B(r)=\ell^2\,\frac{r^2-r_+^2}{r_+^2-r_-^2} \ , \qquad A(r) = B(r) + \ell^2\ . 
\end{equation}
Thus, using \eqref{Poincare} we find that in any patch the metric \eqref{Poincare-line-element} is given by
\begin{equation}
    ds^2 = \frac{\ell^2 r^2}{(r^2 - r_+^2)(r^2-r_-^2)}\ dr^2 + \frac{r_+^2+r_-^2-r^2}{\ell^2} \ dt^2 - \frac{2 r_+ r_-}{\ell}\ dt\, d\phi + r^2 d\phi^2,
\end{equation}
which indeed coincides with the BTZ metric \eqref{BTZ-metric}. In  each of the regions the radius is related to the Poincar\'e coordinates in the same way, \eqref{radius-BTZ}. This formula differs from the analogous expression \eqref{radius-AdS} in AdS$_3$. 

The coordinate systems above are such that in each of the regions the Killing vector field that generates identifications is
\begin{equation}
    \xi' = - \frac{r_+}{\ell}\,(z\,\p_z+\beta\p_\beta +\gamma\p_\gamma) +\frac{r_-}{\ell}\,(\beta\p_\gamma +\gamma\p_\beta) = \partial_\phi\ .
\end{equation}
These expressions are equally valid in $\widetilde{\text{AdS}_3}$ and $\mathcal{M}$. Only in the latter case the coordinate $\phi$ is periodic,
\begin{equation}\label{phi}
   \phi\sim\phi+2\pi n \ .                                       
\end{equation}

\section{$SO(2,2)$ and its representations}

In this appendix, we collect some facts and fix conventions related to the group $SO(2,2)$ and its unitary irreducible representations that are used in the main text. Our main references are \cite{Vilenkin,Bargmann:1946me,Kirillov,Repka,Lang}.

\subsection{The group and its Lie algebra}

The Lie algebra of $SO(2,2)$ is spanned by elements $\{M_{\mu\nu}\}$ that obey the bracket relations
\begin{equation}
    [M_{\mu\nu},M_{\rho\sigma}] = \eta_{\mu\rho} M_{\nu\sigma} + \eta_{\mu\sigma} M_{\rho\nu} + \eta_{\nu\rho} M_{\sigma\mu} + \eta_{\nu\sigma} M_{\mu\rho}\ .
\end{equation}
The group $SO(2,2)$ is locally isomorphic to $SL(2,\mathbb{R})\times SL(2,\mathbb{R})$ and it is in fact the latter group that we will be studying. Its unitary irreducible representations are of the form $V_l\otimes V_r$, where $V_l$ and $V_r$ are unitary irreducibles of $SL(2,\mathbb{R})$. In view of this fact, we will in the remainder of this appendix focus on the left copy of $G=SL(2,\mathbb{R})$ and its representations.

Let $\{H,E_\pm\}$ be a basis for the Lie algebra $\mathfrak{g}=\mathfrak{sl}(2,\mathbb{R})$ in which the bracket relations read
\begin{equation}
    [H, E_+] = E_+, \quad [H, E_-] = - E_-, \quad [E_+,E_-] = 2H\ .
\end{equation}
In the {\it fundamental} two-dimensional representation, these generators might be chosen as 
\begin{equation}\label{defining-representation}
    H=\begin{pmatrix}
      \frac 12 & 0\\ 0 & -\frac 12
     \end{pmatrix},\quad  E_+  = \begin{pmatrix}
      0 & 1\\ 0 & 0
     \end{pmatrix}, \quad E_-=   \begin{pmatrix}
      0 & 0\\ 1 & 0
     \end{pmatrix}\ .
\end{equation}
General elements of $G$ in the fundamental representation will be written as
\begin{equation}\label{generic-element}
 g = \begin{pmatrix}
     \alpha &\beta \\ 
     \gamma &\delta
     \end{pmatrix}\ .
\end{equation}
In the following, we will often make use of the element $w\in G$ known as the Weyl inversion, which is defined by
\begin{equation}
    w = e^{\,\frac\pi2(E_- - E_+)} = \begin{pmatrix} 
    0 & -1\\
    1 & 0\end{pmatrix}.
\end{equation}
The Weyl inversion satisfies
\begin{equation}\label{properties-of-inversion}
   w H w^{-1} = -H,\quad w E_+ w^{-1} = -E_-, \quad w E_- w^{-1} = -E_+\ .
\end{equation}
Notice that, while $\text{Ad}_w\,$ is an involutive automorphism of $\mathfrak{g}$, $w$ itself does not square to the identity of $G$, but instead to the non-trivial element of the centre $Z(G) = \{\pm I\}$.  

The maximal compact subgroup of $G$ is the $SO(2)$ group generated by $M=E_+ - E_-$. One can introduce the complex linear combinations of generators
\begin{equation}
    \tilde H = \frac{i}{2}(E_+ - E_-), \quad \tilde E_+ = \frac12(E_+ + E_- + 2iH), \quad \tilde E_- = \frac12 (E_+ + E_- - 2iH)\,,
\end{equation}
which satisfy the same brackets as $\{H,E_+,E_-\}$. These generators are particularly important for the study of discrete series representations. The Lie algebra $\mathfrak{g}$ has the unique Casimir invariant
\begin{equation}
    C_2 = H^2 + \frac12 \{E_+,E_-\}\ .
\end{equation}
The universal covering group $\widetilde{SL(2,\mathbb{R})}$ of $SL(2,\mathbb{R})$ is an infinitely sheeted cover. It consists of equivalence classes of paths $p_g$ in $SL(2,\mathbb{R})$ which start at the identity, the two paths being equivalent if they end at the same point $g$ and can be continuously deformed into one another. By the group multiplication, a path $p_g$ that starts at $e$ may be mapped into one, $hp_g$ that starts at any other point $h$ in $SL(2,\mathbb{R})$, and ends at $hg$. Using this, multiplication in $\widetilde{SL(2,\mathbb{R})}$ is defined through concatenation of paths
\begin{equation}
    [p_{g_1}] [p_{g_2}] := \left[p_{g_1}\cdot (g_1 p_{g_2})\right]\ .
\end{equation}

\subsection{Unitary irreducible representations}

The group $SL(2,\mathbb{R})$ has three series of unitary irreducible representations -- the principal, discrete and complementary series. Below we give more details about the first two of these, which are the only ones used in the main text.

\subsubsection{Principal series representations}

Given any complex number $\,\tau\in\mathbb{C}$ and $\,\varepsilon\in\{0,1/2\}$ one defines a non-unitary principal series representation $\,T_{\tau,\varepsilon}$ of $\,G$ on the Hilbert space $\mathcal{H} = L^2(\mathbb{R})$ with the standard inner product
\begin{equation}\label{inner-product-principal-series}
    \langle f_1, f_2 \rangle = \int\limits_{-\infty}^\infty dx\ \overline{f_1(x)} f_2(x)\ .
\end{equation}
The action of an element $g$ as in \eqref{generic-element} on a function $f$ is defined by
\begin{equation}
    \left(T_{\tau,\varepsilon}(g)f\right)(x) = \text{sgn}\left(\beta x +\delta\right)^{2\varepsilon}|\beta x+\delta|^{2\tau} f\left(\frac{\alpha x+\gamma}{\beta x+\delta}\right)\ .
\end{equation}
The representation is unitary with respect to the above inner product if and only if $\tau$ is of the form $\tau = -1/2 + i\mathbb{R}$. In such cases $\,T_{\tau,\varepsilon}$ is said to belong to the unitary principal series. Two representations $T_{\tau_1,\varepsilon_1}$ and $T_{\tau_2,\varepsilon_2}$ with non-integer $\tau_i$ are equivalent if and only if $\varepsilon_1 = \varepsilon_2$ and $\tau_2\in\{\tau_1,-1-\tau_1\}$. Consequently, if we write $\tau = -1/2 + i\rho$, the unitary principal series runs over $\,\varepsilon=0,1/2$ and $\rho\in(0,\infty)$.

In representations $\,T_{\tau,\varepsilon}$ the generators act as differential operators
\begin{equation}
    E_- = \partial_x, \quad H = x\partial_x - \tau, \quad E_+ = -x^2 \partial_x + 2\tau x\ .
\end{equation}
In the unitary principal series, each of the operators $\,\{iH,iE_+,iE_-\}\,$ has the spectrum $(-\infty,\infty)$. These are not highest weight representations. Finally, the value of the quadratic in $\,T_{\tau,\varepsilon}$ is
\begin{equation}
    C_2(T_{\tau,\varepsilon}) = \tau(\tau+1) = - \left(\rho^2 + \frac14\right)\ .
\end{equation}

\subsubsection{Discrete series representations}

Discrete series is defined as the set of those unitary irreducible representations of $G$ whose matrix elements are square integrable functions on the group. The group $SL(2,\mathbb{R})$ has two sets of discrete series representations, $T_l^-$ and $T_l^+$. Here, $l$ is a half integer and in the first case $l\leq-1$, while in the second $l\geq1$. We will describe representations $\,T^-_l$, the other ones being similar.

The carrier space of $T_l^-$ is that of analytic functions in the upper half plane, square integrable with respect to the inner product
\begin{equation}\label{inner-product-discrete-series}
 (F_1,F_2) = \frac{1}{2\Gamma(-2l-1)}\int\limits_{-\infty}^\infty dx\int\limits_0^\infty dy \, y^{-2l-2} \,\overline{F_1(z)} F_2(z) \ .
\end{equation}
This space will be denoted by $\mathcal{H}_l$. The action of the $SL(2)$-matrix \eqref{generic-element} on a function $F$ is defined as
\begin{equation}
    \big(T_l^-(g) F\big)(z) = (\beta z +\delta)^{2l} \, F\Big( \frac{\alpha z +\gamma}{\beta z+\delta}\Big)\ .
\end{equation}
Therefore, the generators are represented by differential operators
\begin{equation}
   T_l^-(H) = z\partial_z-l ,\quad T_l^-(E_+) = -z^2\partial_z + 2 l z,  \quad T_l^-(E_-) = \partial_z\ .
\end{equation}
Contrary to the principal series, representations $\mathcal{H}_l\,$ have lowest-weight vectors. To exhibit the lowest-weight structure, one considers the generator $E_+ - E_-$ of the $SO(2)$ subgroup of $G$. In order for the eigenfunctions of $\tilde H$ to be holomorphic at $i$, its eigenvalues must take the form $-l+\mathbb{N}_0$. The eigenfunctions are
\begin{equation}\label{discrete-basis}
    \Psi_n(z) = (z-i)^n (z+i)^{2l-n}, \quad n = 0,1,\dots\ .
\end{equation}
These functions are square integrable with respect to \eqref{inner-product-discrete-series}, \cite{Lang}, and the Hilbert space decomposes as the sum of one-dimensional eigenspaces of $\tilde H$, $\mathcal{H}_l = \oplus \langle\Psi_n\rangle$. The representation has the lowest weight vector $\Psi_0 = (z+i)^{2l}$ that satisfies
\begin{equation}\label{lowest-weight}
    \tilde H \Psi_0 = -l \Psi_0, \quad \tilde E_- \Psi_0 = 0\,,
\end{equation}
and a basis for $\mathcal{H}_l$ is obtained by successive applications of $\tilde E_+\,$ to it.

As in the principal series, the spectrum of $\,iH$ in $\,T_l^-$ is $(-\infty,\infty)$. However, the spectra of $iE_-$ and $iE_+$ are $(-\infty,0)$ and $(0,\infty)$, respectively. The quadratic Casimir is positive in discrete series representations and assumes the value $\, C_2(T_l^-)=l(l+1)$.

In the main text, we have used the decomposition of the endomorphism algebra $\text{End}(\mathcal{H}_l)$ into irreducible components. This result is derived from two facts. Firstly, the dual (contragredient) representation of $\,T_l^-$ is $\,T_{-l}^+$. Therefore, the question is equivalent to that of decomposing $\,T_l^-\otimes T_{-l}^+$, which may be found in \cite{Repka}. The derivation of \cite{Repka} passes through an intermediate step which realises representations on the two sides on the space of functions on $\mathbb{H}^2$. Therefore, one has the sequence of module isomorphisms
\begin{equation}
    \text{End}(\mathcal{H}_l) \cong T_l^- \otimes T_{-l}^+ \cong L^2(\mathbb{H}^2,d\mu) \cong \text{Ind}_K^G 1 \cong \int\limits_0^\infty d\rho\ T_{i\rho-1/2,0}\ .
\end{equation}
In this equation, the measure $d\mu$ is the standard one on the hyperbolic space, $d\mu = y^{-2}dx\,dy$. We have displayed it explicitly because the same space with a different measure can appear as the carrier space for other kinds of tensor products of the form $\,T_m^-\otimes T_{-n}^+$.

\subsubsection{Discrete series in the Fourier space}

Another realisation of discrete series representation that we found useful is based on the fact that functions that are holomorphic in the upper half-plane are Fourier transforms of functions defined on the semi-axis $(0,\infty)$, 
\begin{equation}
    F(z) = \int\limits_0^\infty dx\ e^{ixz} \hat F(x)\ .
\end{equation}
Under the Fourier transform, the inner product $(\ref{inner-product-discrete-series})$ becomes
\begin{equation}\label{inner-produt-Fourier}
    (\hat F_1,\hat F_2) = 2^{2l+1}\pi \int\limits_0^\infty dx\ x^{2l+1} \overline{\hat F_1(x)} \hat F_2(x)\ .
\end{equation}
In Fourier space, generators of $SL(2,\mathbb{R})$ are no longer represented by first-order differential operators. Explicit expressions may be found in \cite{Vilenkin}. We will only require the following two results
\begin{equation}\label{generators-Fourier}
    H = -x\partial_x - l - 1, \quad\quad E_- = ix\ .
\end{equation}
In the main text, we use the representation conjugated by the Weyl inversion $w$, which has
\begin{equation}\label{generators-Fourier-w}
    H = x\partial_x + l + 1, \quad\quad E_+ = - ix\ .
\end{equation}
From here it is also clear that $\text{Spec}(iE_-)=(-\infty,0)$, with eigenfunctions $\hat f_\lambda(x) = \delta(x+\lambda)$.

\section{Below-horizon states}

In this appendix we give the eigenfunctions corresponding to the negative eigenvalues of the regularised operator $B$.

Let us first consider the potential $V^\ast $ and its the bound states $\,h_\lambda^\ast$, $\,\lambda^2=-\kappa^2<0\,$,
\begin{equation}
     V^\ast(\chi) =\left\{ \begin{array}{ll}
      \infty \, , \quad  &  \chi\in (0, \chi_0)  \\[8pt]
     - \left(c^2+\dfrac 14\right)\,\dfrac{1 }{\chi^2} \, , \ \    &  \chi\geq\chi_0 \ 
    \end{array}   \right.   \\[2pt]
\end{equation}
where as before,
\begin{equation}
     4\kappa_0^2= \frac{4 r_+^2}{r_+^2-r_-^2}  = (c^2+\frac 14)\, \frac{1}{\chi_0^2}\ ,
     \quad c=c(n)=\frac{n\ell}{r_-} \ . \\[2pt]
\end{equation}
We will discuss solutions corresponding to a fixed $n$. The eigenfunctions are given by
\begin{equation}
    h_\lambda^\ast(\chi) = \left\{ \begin{array}{ll}
       0\,, &  \chi\in (0, \chi_0)\\[8pt]
        D \sqrt{\chi} \, \, K_{ic}(2\kappa\chi)\,, \quad  &  \chi\geq\chi_0 \ 
\end{array}    \right.
\end{equation}
for some constant $D$ and they obey the continuity condition
\begin{equation}           \label{continuity}
K_{ic}(2\kappa\chi_0)=0\ .
\end{equation}
The asymptotic behaviour of modified Bessel functions near zero is
 \begin{equation}\label{KBessel}
     K_{ic}(2\kappa\chi_0) =  -\sqrt{\frac{\pi}{c\sinh\pi c}\, }\, \sin\Big(c\log(\kappa\chi_0)-\gamma_c\Big) +O(\kappa^2)\,,       
 \end{equation}
where $\gamma_c\,$ is the phase of $\Gamma(1+ic)$. These asymptotics imply that there are infinitely many solutions $\,\kappa_N$ of \eqref{continuity} when $\kappa$ approaches 0,
\begin{equation}\label{asymptotic-energies}
     \chi_0\kappa_N \approx e^{\, \frac{\gamma_c}{c}-\frac{N\pi}{c}}\ .
\end{equation}

The other regularised potential  $\, V^\star$ can be analysed along similar lines and leads to results which are qualitatively the same. The bound state solutions for 
\begin{equation}
    V^\star(\chi) =\left\{ \begin{array}{ll}
      -\, \dfrac{4 r_+^2}{r_+^2-r_-^2} \, , \quad  &  \chi\in (0, \chi_0)  \\[12pt]
     - \Big(c^2+\dfrac 14\Big)\,\dfrac{1 }{\chi^2} \,, \ \    &  \chi\geq\chi_0\,,
    \end{array}   \right. 
\end{equation}
are of the form
\begin{equation}                       
    h_\lambda^\star(\chi) =\left\{ \begin{array}{lll}
     A\cos k\chi + B\sin k\chi \, , \quad   & k^2  = 4(\kappa_0^2-\kappa^2)  ,\qquad &  \chi\in (0, \chi_0) \,    \\[8pt]
  D\sqrt{\chi}\, K_{ic}(2\kappa\chi)   \, , &\,  \qquad  &  \chi>\chi_0 \,,
    \end{array}   \right.  \\[4pt]
\end{equation}
where $A$, $B$ and $D$ are constants. Continuity conditions at $\chi_0$ give the quantisation of $\kappa$. Denoting $\, \tan\Upsilon =A/B$, we find that the allowed values of $\,\kappa$ are solutions to the equation
\begin{equation}
   2 \sqrt{c^2+\frac 14 -(2\kappa\chi_0)^2}\  \cot \Big( \Upsilon + \sqrt{c^2+\frac 14 -(2\kappa\chi_0)^2} \ \Big)
    = 1+4\,\kappa\chi_0  \, \frac{\, K'_{ic}(2\kappa\chi_0)\, }{\, K_{ic}(2\kappa\chi_0)\, } \,\ .
\end{equation}
Expanding this equation for small values of $\kappa$ and using \eqref{KBessel}, we find that the solutions are
\begin{equation}
    c\,\log (\chi_0\kappa_N) = -N\pi +C \, ,    \qquad N\in \mathbb{N}\, ,
\end{equation}
where $C$ is a  constant depending on parameters $\,c$, $ \chi_0$ and $ \Upsilon\,$.

\end{document}